\documentclass[a4paper,11pt]{article} 
\usepackage[utf8]{inputenc}
\usepackage[greek,english]{babel}
\languageattribute{greek}{ancient}
\usepackage{amsfonts,amsmath}
\usepackage{latexsym}
\usepackage{commath}
\usepackage{fontawesome}
\usepackage[makeroom]{cancel}
\usepackage{amssymb}
\usepackage{amsthm}
\usepackage{graphicx}
\usepackage{dsfont}
\usepackage{scalerel}
\usepackage[font=footnotesize]{caption}
\usepackage{bigints}
\usepackage[T2A,T1]{fontenc}
\usepackage{physics}
\usepackage[dvipsnames]{xcolor}
\usepackage{float}
\usepackage{authblk}
\usepackage{relsize,exscale}
\usepackage{mathtools}
\usepackage{xparse}
\usepackage{siunitx}
\usepackage{array}
\usepackage{etoolbox}
\usepackage{tensor}
\usepackage{nicefrac}

\usepackage{bm}
\usepackage[numbers,sort&compress]{natbib}
\usepackage{tocloft}  
\usepackage{spalign}
\usepackage{url}
\usepackage{microtype}
\usepackage{paralist}
\usepackage{wrapfig}
\usepackage{esint}
\usepackage{mathdots}
\usepackage{ffff}
\usepackage{indentfirst}
\usepackage{titlesec}
\usepackage{hyperref}
\numberwithin{equation}{section}
\DeclareRobustCommand{\varlambda}{\text{\usefont{OML}{txmi}{m}{it}\symbol{"15}}}
\DeclareRobustCommand{\vargamma}{\text{\usefont{OML}{txmi}{m}{it}\symbol{"0D}}}

\makeatletter
\newcommand*{\textoverline}[1]{$\overline{\hbox{#1}}\m@th$}
\makeatother
\DeclareFontEncoding{LS1}{}{}
\DeclareFontSubstitution{LS1}{stix}{m}{n}
\DeclareSymbolFont{stixletters}{LS1}{stix}{m}{it}
\DeclareMathAccent{\cev}{\mathord}{stixletters}{"91}
\DeclareMathAccent{\vec}{\mathord}{stixletters}{"92}
\DeclareMathAccent{\vecev}{\mathord}{stixletters}{"95}
\titleformat{\section}[block]{\large\bfseries\centering}{\thesection}{1em}{} 
\newcommand*\pFqskip{8mu}
\catcode`,\active
\newcommand*\pFq{\begingroup
	\catcode`\,\active
	\def ,{\mskip\pFqskip\relax}%
	\dopFq
}
\catcode`\,12
\def\dopFq#1#2#3#4#5{%
	{}_{#1}F_{#2}\biggl[\genfrac..{0pt}{}{#3}{#4};#5\biggr]%
	\endgroup
}
\catcode`,\active
\newcommand*\tpFq{\begingroup
	\catcode`\,\active
	\def ,{\mskip\pFqskip\relax}%
	\dotpFq
}
\catcode`\,12

\def\dotpFq#1#2#3#4#5{%
	{}_{#1}\widetilde{F}_{#2}\biggl[\genfrac..{0pt}{}{#3}{#4};#5\biggr]%
	\endgroup
}

\colorlet{myfilecolor}{violet}
\colorlet{myurlcolor}{Aquamarine}
\colorlet{mylinkcolor}{YellowOrange}
\definecolor{skobeloff}{rgb}{0.0, 0.48, 0.45}
\definecolor{hi}{RGB}{16, 9, 159}

\makeatletter
\newcommand{\megLeftrightarrow}[2][]{\ext@arrow 0099\Leftrightarrowfill@{#1}{#2}}
\makeatother

\DeclareFontFamily{U}{matha}{\hyphenchar\font45}
\DeclareFontShape{U}{matha}{m}{n}{
	<5> <6> <7> <8> <9> <10> gen * matha
	<10.95> matha10 <12> <14.4> <17.28> <20.74> <24.88> matha12
}{}
\DeclareSymbolFont{matha}{U}{matha}{m}{n}
\DeclareMathSymbol{\updownharpoons}{3}{matha}{"EA}
\DeclareMathSymbol{\downupharpoons}{3}{matha}{"EB}
\DeclareMathSymbol{\upharpoonleft}{3}{matha}{"E4}
\DeclareMathSymbol{\downharpoonright}{3}{matha}{"E7}
\DeclareFontFamily{U}{mathb}{\hyphenchar\font45}
\DeclareFontShape{U}{mathb}{m}{n}{
	<5> <6> <7> <8> <9> <10> gen * mathb
	<10.95> mathb10 <12> <14.4> <17.28> <20.74> <24.88> mathb12
}{}
\DeclareSymbolFont{mathb}{U}{mathb}{m}{n}
\DeclareMathSymbol{\upupharpoons}{3}{mathb}{"DA}
\DeclareMathSymbol{\downdownharpoons}{3}{mathb}{"DB}

\hypersetup{
	colorlinks=true,
	linkcolor=hi,
	filecolor=hi,      
	urlcolor=hi,
	citecolor=hi,
	linktoc=page,
	pdfnewwindow=true,
}
\makeatletter
\renewcommand*{\@fnsymbol}[1]{%
	\ensuremath{%
		\ifcase#1
		\or
		\natural 
		\or
		\sharp   
		\else
		\@arabic{\numexpr#1 -2\relax} 
		\fi
	}%
}
\makeatother

\newcommand{\al}{\alpha}
\newcommand{\be}{\beta}
\newcommand{\g}{\gamma}
\newcommand{\de}{\delta}
\newcommand{\e}{\epsilon}

\newcommand{\ka}{\kappa}
\newcommand{\la}{\lambda}
\newcommand{\m}{\mu}
\newcommand{\n}{\nu}

\newcommand{\ta}{\tau}

\newcommand{\La}{{\mathcal{L}}}

\DeclareSymbolFont{symbolstx}{OMS}{txsy}{m}{n}
\SetSymbolFont{symbolstx}{bold}{OMS}{txsy}{bx}{n}
\DeclareSymbolFontAlphabet{\mathcall}{symbolstx}

\def\xcpm#1#2#3{\mathbin{\ooalign{%
			\raise #2\hbox{\pdfliteral{1 0 0 rg}$#1+$\pdfliteral{0 g}}\cr
			\lower #3\hbox{\pdfliteral{0 0 1 rg}$#1-$\pdfliteral{0 g}}%
}}}
\title{\textbf{Massive gravity applications for \(T\overline{T}\) deformations} }
\author{Alexia Nix\textsuperscript{$\flat$,}\thanks{alexianix@hi.is \href{mailto:alexianix@hi.is}{\faEnvelopeO}}\phantom{,} \& Evangelos Tsolakidis\textsuperscript{$\flat$,}\thanks{evangelos@hi.is \href{mailto:evangelos@hi.is}{\faEnvelopeO}}\vspace{0.2cm}}
\affil{\textsuperscript{$\flat$}\emph{Science Institute, University of Iceland,}\\ \emph{Dunhaga 3, 107 Reykjavík,} \\ \emph{Iceland}\vspace{-0.8cm}}
\date{}
\begin{document} 
	\maketitle
	\begin{abstract}
		\normalsize
		\noindent We employ the massive gravity approach to stress-tensor deformations in a variety of scenarios, obtaining novel results and establishing new connections. Starting with perturbation theory, we show that the addition of \(\tr T+\Lambda_{2}\) to \(T\overline{T}\) can be recovered and we construct the deformed action of an interacting non-abelian spin-1 along with spin-1/2 seed model, extending previous findings. As a result, a set of algebraic properties for certain hypergeometric functions is derived, allowing us to initiate the algebraic study of special functions directly via stress-tensor deformations and massive gravity. Moreover, we sharpen the connection between the trace-flow equation and the local renormalization group in any dimension. In \(d>2\), the usual initial condition for the coupling leads to a potential known as ghost-free, minimal massive gravity. Upon expansion around the reference background, we retrieve Fierz-Pauli at leading order, matching the random geometry and holographic approaches. At the same time, we demonstrate that a change of coordinates interpretation is possible for the corresponding operator, which we verify with a simple example. Finally, we study the family of \((\tr T)^{n}\) deformations advancing earlier work, and illustrate how the massive gravity description of non-linear electrodynamics can be incorporated in our framework.
	\end{abstract}
	\thispagestyle{empty}
	\newpage
\beforetochook\hrule
\tableofcontents
\afterTocSpace
\hrule
\afterTocRuleSpace
\flushbottom
\newpage
\section{Introduction}
This work explores various novel gravitational aspects of the \(T\overline{T}\) program and its extensions. It is considered standard practice in QFT to discard operators with mass dimension greater than the one of the ambient spacetime under the RG flow process, as they only become important in the UV or treat the deformed theory as an EFT. Over the past decade however, considerable effort has been dedicated towards understanding the physics of deformations with such peculiar properties in a more systematic way. In more detail, most results in this research direction are derived in two dimensions and involve the well-known by now \(T\overline{T}\) operator \cite{Zamolodchikov:2004ce} i.e. the determinant of the stress-tensor of a theory. The underlying motivation comes from multiple, and sometimes orthogonal viewpoints, as recent discoveries continue to expose the field’s rich structure, something we also aim to highlight with this paper.

 Historically, these deformations were considered as another tool one can use to map out the space of all possible QFT's  \cite{Smirnov:2016lqw}. On the other hand, interesting connections to string theory and integrability were noted in \cite{Cavaglia:2016oda}  based on earlier results \cite{Caselle:2013dra}. In short, a concrete classical connection between the Nambu-Goto action and the \(T\overline{T}\) deformation of a CFT was established, and the corresponding energy spectrum (on a cylinder) was found to precisely match the one of closed strings in a constant \(B\)-field under certain identifications. Soon after, the discovery of the holographic interpretation of the \(T\overline{T}\) operator revealed another avatar of this program where the deformed CFT could be thought of living on a radial slice in the bulk of AdS\(_3\), at depth inversely proportional to the deformation parameter \(\la\) \cite{McGough:2016lol} (see \cite{Guica:2019nzm} for a more recent approach). These results were later extended to arbitrary dimensions \cite{Taylor:2018xcy,Hartman:2018tkw}, providing in parallel a higher-dimensional operator which could serve as the analogue of \(T\overline{T}\). Around the same time, the single trace version of this deformation was proposed \cite{Giveon:2017nie, Giveon:2017myj}, where interesting connections to little string theory were noted. In another intriguing formulation, flat JT gravity coupled to a QFT was found to be equivalent to the \(T\overline{T}\) deformation of that QFT in flat spacetime  \cite{Dubovsky:2017cnj,Dubovsky:2018bmo}, with the uplift to a curved background appearing in \cite{Tolley:2019nmm, Mazenc:2019cfg}. The gravitational dynamics in that case are governed by a particular type of ghost-free massive gravity, which was already well-studied in various dimensions, see e.g. \cite{deRham:2010kj, deRham:2014zqa,Hassan:2011vm,Hassan:2011zd}. A closely related approach considers a spin-2 Hubbard-Stratonovich transformation as the source of the deformation \cite{Cardy:2018sdv}, which can be matched to the linearized massive gravity description obtained by the leading order expansion in \(\la\).
 
 The aforementioned topics focus mainly on defining reformulations of \(T\overline{T}\), but there are some other interesting extensions we would like to mention.\footnote{The list that follows is by no means exhaustive,  for a pedagogical introduction or a recent review one may refer to \cite{Jiang:2019epa, He:2025ppz}.} A nice framework for obtaining the classically deformed action was constructed in \cite{Bonelli:2018kik} and \cite{Frolov:2019nrr}. There also exists a family of non-relativistic deformations which usually involve the current of a \(U(1)\) symmetry \cite{Guica:2017lia, Frolov:2019xzi}. In the same spirit, several connections to non-Lorentzian geometry have been pointed out \cite{Blair:2020ops,Hansen:2020hrs,Blair:2024aqz}. Furthermore, deformed correlators have been studied, see for example \cite{Kraus:2018xrn,Cardy:2019qao,Aharony:2023dod}.  Other formulations of geometrical origin include \cite{Conti:2018tca, Conti:2022egv}, which are also valid in higher dimensions. It is worth noting that there are by now a number of generalizations of stress-tensor deformations, with the most prominent example being root-\(T\overline{T}\) \cite{Rodriguez:2021tcz,Ferko:2022cix,Babaei-Aghbolagh:2022leo}, a four-dimensional version of which \cite{Babaei-Aghbolagh:2022uij} was found to be related to a non-linear conformal extension of electrodynamics, known as ModMax theory \cite{Bandos:2020jsw}.
 
With this paper, we aim to further establish our previously noted connection between massive gravity theories and stress-tensor deformations. The central relation that governs the dynamics of all our findings is 
\begin{gather}
	\pdv{S_{\la}}{\la}=\mathcal{O}\,, \label{floweq}
\end{gather}
which will be referred to as the flow equation. The parameter \(\la\) will serve as the coupling of the operator \(\mathcal{O}\) which implicitly depends on it, and \(S_{\la}\) is the deformed action. As briefly mentioned above, our results have an interdisciplinary character. Remarkably, employing our method allowed us to derive various non-trivial algebraic identities for certain hypergeometric functions, extending a collection known as Schwarz's list \cite{Schwarz1873, Brennan:2019azg}. On the purely gravitational side of the story, we find that our framework very naturally connects and extends elements from the holographic renormalization of AdS\(_{d+1}\) \cite{Taylor:2018xcy, Hartman:2018tkw,deHaro:2000vlm,Balasubramanian:1999re} with the very well-known ghost-free minimal massive gravity theory \cite{Hassan:2011vm,Hassan:2011zd} in \(d\) dimensions. At the same time, we readily demonstrate the existence of dynamical coordinates as a tool for obtaining either deformed actions or saddles for the special class of operators we proposed, extending the two-dimensional formalism of \cite{Conti:2018tca}. We also show how the massive gravity interpretation of the Born-Infeld theory \cite{Floss:2023nod} fits very naturally in our description, and as a result we provide a concrete way of deforming any Weyl invariant theory in four dimensions. While these constitute our main contributions, we further test our structure in other contexts, verifying its robustness in every case.
 
The remainder of this paper is structured as follows. In Section \ref{sec2} we review the massive gravity techniques used to analyse stress-tensor deformations in $d$ dimensions and supplement the existing literature on this topic with new universal results. We continue our analysis in Section \ref{sec3} with applications to two- and one-dimensional theories, extending previous findings. In parallel, using the massive gravity approach we propose an algorithm that can potentially determine whether a given special function can be expressed as a solution in radicals, and we comment on the relation between the local RG flow and trace-flow equations. In Section \ref{sec4}, we study a variety of deformed theories and their corresponding gravitational dressings, gaining novel insights from both perspectives. We collect our main results in Section \ref{disc}, discuss possible future directions, and close with a few comments on deformations for pure gravity, highlighting important caveats stemming from the first versus second order formalism. Finally, in Appendix \ref{conv} we summarize useful properties of the elementary symmetric polynomials.
\section{Deformations via massive gravity} \label{sec2} 
In this section we will briefly review the methodology developed in \cite{Tsolakidis:2024wut}, highlighting and explaining the relevant results for the present work. Let us begin with the ansatz for the deformed action which reads
\begin{gather}
	S_{\la}[\tensor{e}{_{\m}^{a}},\tensor{f}{_{\m}^{a}},\mathfrak{g}_{i},\Phi]=\int\dd[d]{x}f\hspace{0.05cm}G(\la,\mathfrak{g}_{i},y_{1},\dots,y_{d})+S_{0}[\tensor{e}{_{\m}^{a}},\mathfrak{g}_{i},\Phi]\,, \label{sol1}
\end{gather}
where \(\Phi\) represents matter, \(\mathfrak{g}_{i}\) couplings of the undeformed (seed) action \(S_{0}\) and \(\tensor{e}{_{\m}^{a}}/\tensor{f}{_{\m}^{a}}\) will be the auxiliary/reference background vielbein.\footnote{Please note that the corresponding metrics will be given by  \(\tensor{g}{_{\m}_{\n}}=\tensor{e}{_{\m}^{a}}\tensor{\eta}{_{a}_{b}}\tensor{e}{_{\n}^{b}}\) and \(\tensor{\g}{_{\m}_{\n}}=\tensor{f}{_{\m}^{a}}\tensor{\eta}{_{a}_{b}}\tensor{f}{_{\n}^{b}}\). We will use the mostly plus Lorentzian signature.} The variables \(y_{i}\) are defined as \(y_{i}\coloneqq\tensor{\qty(Y_{i})}{^{\m}_{\m}}=\tr(Y_{i})\) where \(\tensor{\qty(Y_{i})}{^{\m}_{\n}}\coloneqq\tensor{f}{^{\m}_{a_{1}}}\tensor{e}{_{\la_{1}}^{a_{1}}}\tensor{f}{^{\la_{1}}_{a_{2}}}\tensor{e}{_{\la_{2}}^{a_{2}}}\cdots\tensor{f}{^{\la_{i-1}}_{a_{i}}}\tensor{e}{_{\n}^{a_{i}}}\) and we will usually refer to the first term of \eqref{sol1} as \(S_{G}\) or the massive gravity dressing, which is usually determined uniquely once the operator \(\mathcal{O}\) is fixed. From there, taking the derivative with respect to the deformation parameter leads to
\begin{gather}
	\dv{S_{\la}}{\la}=\int\dd[d]{x}f\hspace{0.05cm}\partial_{\la}G(\la,\mathfrak{g}_{i},y_{1},\dots,y_{d})+\int\dd[d]{x}\frac{\de S_{\la}}{\de \tensor{e}{_{\m}^{a}}}\dv{\tensor{e}{_{\m}^{a}}}{\la}\,. \label{der1}
\end{gather}
 Notice that we took the total derivative whereas the definition \eqref{floweq} involves the partial. For an honest deformed theory e.g. the Nambu-Goto string in two dimensions coming from the \(T\overline{T}\) deformation of a free scalar boson \cite{Cavaglia:2016oda}, this does not make any difference whatsoever. However, picking \eqref{sol1} as the solution the flow equation introduces a fictitious degree of freedom that is the auxiliary\footnote{The reference background \(\tensor{f}{_{\m}^{a}}\) is the fixed geometry that our theory lives on, thus independent of \(\la\).} background \(\tensor{e}{_{\m}^{a}}\), which needs to be eliminated one way or another. Since the fields \(\Phi\), the background \(\tensor{f}{_{\m}^{a}}\) and the other couplings \(\mathfrak{g}_{i}\) are all independent of \(\la\) by construction, the total derivative is only able to ``see'' \(\tensor{e}{_{\m}^{a}}\) as the equation above suggests. This is not necessary, other approaches may also apply here, however it will allow us to completely fix the auxiliary vielbein in a very convenient way. 
 
  On the other hand, considering that the generic operator is given by 
 \begin{gather}
 	\mathcal{O}=\int\dd[d]{x}f\hspace{0.05cm}\mathcall{O}(\la,\mathfrak{g}_{i},\tensor{T}{^{\m}_{a}})\,, \qquad \tensor{T}{^{\m}_{a}}\coloneqq\frac{1}{f}\frac{\de S_{\la}}{\de \tensor{f}{_{\m}^{a}}}\,,
 \end{gather}
 the flow equation \eqref{floweq} can be brought in the following form
 \begin{gather}
 	\partial_{\la}G(\la,\mathfrak{g}_{i},y_{1},\dots,y_{d})=\mathcall{O}(\la,\mathfrak{g}_{i},w_{1},\dots,w_{d})\,, \label{pde}
 \end{gather}
i.e. a first order, \((d+1)\)-dimensional non-linear partial differential equation,\footnote{The index \(i\) for the variables \(y_{i},w_{i}\) runs from \(1\) to \(d\) due to the Cayley-Hamilton theorem. Generally, for any \(y_{k}\) with \(k>d\) we can use the theorem to recast \(y_{k}\) as function of \((y_{1},\dots,y_{d})\).} where \(w_{i}\coloneqq\tensor{\qty(W_{i})}{^{\m}_{\m}}=\tr(W_{i})\) and \(\tensor{\qty(W_{i})}{^{\m}_{\n}}\coloneqq\tensor{T}{^{\m}_{a_{1}}}\tensor{f}{_{\la_{1}}^{a_{1}}}\tensor{T}{^{\la_{1}}_{a_{2}}}\tensor{f}{_{\la_{2}}^{a_{2}}}\cdots\tensor{T}{^{\la_{i-1}}_{a_{i}}}\tensor{f}{_{\n}^{a_{i}}}\). The   above, uniquely determines \(G\) given that \(w_{i}\) can be written as \cite{Tsolakidis:2024wut}
\begin{gather}
	w_{i}=dG^{i}+\sum_{k=1}^{i}(-1)^{k}G^{i-k}\binom{i}{k}\sum_{j_{1},\dots,j_{k}=1}^{d}\qty[\prod_{n=1}^{k}j_{n}\pdv{G}{y_{j_{n}}}]y_{j_{1}+\cdots+j_{k}}\,. \label{wa}
\end{gather}
We notice that \eqref{pde} essentially forces \(G\) into containing all the information regarding the deformation \(\mathcall{O}\). In other words, \(S_{G}\) in \eqref{sol1} is a dressing that is exclusively responsible for the deformation, leaving \(S_{0}\) untouched. But this is precisely one of the most important properties of stress-tensor deformations; they are valid for any seed theory! At the same time, as \(\la \rightarrow 0\) one recovers only the seed action which now lives on the reference background i.e.
\begin{gather}
	\tensor{e}{_{\m}^{a}}\Big{|}_{\la\rightarrow 0}=\tensor{f}{_{\m}^{a}}\,, \qquad	G(0,\mathfrak{g}_{i},d,\dots,d)=0\,. \label{inco}
\end{gather}   

The final assumption of this approach is that we want to work with a deformed action that cannot distinguish between the total and the partial derivative with respect to \(\la\), just like the Nambu-Goto case we discussed above. This leads to the following constraint
\begin{gather}
	\qty(\dv{}{\la}-\pdv{}{\la})S_{\la}=0=\int\dd[d]{x}\frac{\de S_{\la}}{\de \tensor{e}{_{\m}^{a}}}\dv{\tensor{e}{_{\m}^{a}}}{\la}\,, \label{condtt}
\end{gather}  
where \eqref{der1} and \eqref{pde} were used. From there the right hand side of the equation above can be valid in many ways,\footnote{We exclude the possibility of \(\dd{\tensor{e}{_{\m}^{a}}}/\dd{\la}=0\) as this choice leads to an arbitrary auxiliary background, albeit independent of \(\la\). \label{foote}} however one should recall that \(\tensor{e}{_{\m}^{a}}\) contains \(d^{2}\) components.\footnote{We do not make any gauge choice for now.} Therefore, in order to uniquely fix it we are guided to the following condition
\begin{gather}
	\frac{\de S_{\la}}{\de \tensor{e}{_{\m}^{a}}}=0\,, \label{intout}
\end{gather}
essentially integrating \(\tensor{e}{_{\m}^{a}}\) out, rightfully justifying its auxiliary nature and title. Using the above, one can find the solution\footnote{In subsection \ref{sectr}, we will find that this constraint does not have any solutions in some cases, therefore we will have to resort to the weaker condition \eqref{condtt}.}  \(\tensor{\bar{e}}{_{\mu}^{a}}(\tensor{f}{_{\m}^{a}},\lambda,\mathfrak{g}_{i},\Phi)\) and also relate \(S_{G}\), the deformed and undeformed stress-tensors like so
\begin{gather}
	f\tensor{\bar{T}}{^{\m}_{a}}\tensor{f}{_{\n}^{a}}=f\bar{G}(\bar{y}_{1},\dots,\bar{y}_{d})\tensor{\de}{^{\mu}_{\n}}+\bar{e}\tensor{(\bar{T}_{0})}{^{\m}_{a}}\tensor{\bar{e}}{_{\n}^{a}}\,, \qquad  	\frac{\de \bar{S}_{\la}}{\de \tensor{f}{_{\m}^{a}}}=	\frac{\de \bar{S}_{G}}{\de \tensor{f}{_{\m}^{a}}}\,,
\end{gather}
where \(\tensor{(T_{0})}{^{\m}_{a}}\) is the undeformed stress-tensor given by the variation with respect to the auxiliary background and the bar indicates which terms are evaluated on \(\tensor{\bar{e}}{_{\mu}^{a}}\). Another important consequence of \eqref{intout} reads as follows
\begin{gather}
	\frac{\de \bar{S}_{\la}}{\de (\star)}=\frac{\de \bar{S}_{0}}{\de (\star)}\,, \label{eom fields}
\end{gather}
that is, on-shell for  \(\tensor{e}{_\mu^a}\), any variation of the deformed action with respect to anything \((\star)\) that is not a background vielbein is equal to the same variation but of the undeformed action. This essentially illustrates that the deformed and undeformed matter equations of motion are the same on \(\tensor{\bar{e}}{_{\mu}^{a}}\)! Furthermore, keep in mind that the first equation of \eqref{inco} should naturally drop out from the solution of \eqref{intout}. We have to also mention that there can exist cases such that \eqref{pde} holds only when \eqref{intout} is true, complicating significantly our attempt to specify \(G\). Interestingly, we will discover in subsection \ref{nonlin} that the theory of non-linear electrodynamics falls precisely within this category.  
\subsection{Initial conditions}
Before we move on to the next section, let us provide more insight regarding the initial condition \eqref{inco}. For the simplest case --the one we will consider for the entirety of this work-- where there are no other dimensionful couplings available, the massive gravity density\footnote{From \eqref{sol1} the mass dimension of \(G\) is \(d\).} factorizes\footnote{It is obvious that in the presence of another dimensionful coupling say \(\mu\), there exists a function such that \(\qty[\varphi(\mu,\la)]=M^{0}\), therefore the factorization does not hold as \(\varphi\) can now combine freely with \(y_{i}\).} as follows
\begin{gather}
	G(\la,0,y_{1},\dots,y_{d})=\la^{d/\Delta_{\la}}K(y_{1},\dots,y_{d})\,, \label{gdef}
\end{gather}
 with \(\Delta_{\la}\) being the mass dimension of \(\la\) since \(y_{i}\) is dimensionless by construction. Considering the dimensional analysis above, we conclude that \(\la\) cannot be a marginal parameter. This also means that marginal operators like root-\(T\overline{T}\) do not have a massive gravity description on their own, they instead dress the dimensionless variables \(y_{i}\) in a particular way \cite{Tsolakidis:2024wut,Babaei-Aghbolagh:2024hti}. Looking back to \eqref{inco}, we observe that if \(\Delta_{\la}>0\) then the second equation is trivially satisfied even if we do not set \(y_{i}=d\). This is problematic since in that case, \(G\) is not aware that the limit \(\la \rightarrow 0\) also enforces \(\tensor{e}{_{\m}^{a}}=\tensor{f}{_{\m}^{a}}\). We are left with only one option; \(\Delta_{\la}<0\) which automatically makes the deformation irrelevant. It is quite remarkable that one can extract so many properties about our theory only via dimensional analysis. Returning to the initial condition, we have to be very careful since \(G\) approaches infinity due to the negative \(\Delta_{\la}\), unless \(K\) approaches zero as \(\la^{-\alpha d/\Delta_{\la}}\), where \(\alpha\) is greater than one. This implies that \(\tensor{e}{_{\m}^{a}}\) has to be an implicit function of \(\la\) that respects \eqref{inco}. Expanding now \(G\) around \(\la\rightarrow 0\) we obtain the following
 \begin{gather}
 	G(\la,0,y_{1},\dots,y_{d})=\sum_{n=0}^{\infty}\lambda^{n+d/\Delta_{\la}}\frac{1}{n!}\dv[n]{K(y_{1},\dots,y_{d})}{\la}\Bigg{|}_{\lambda\rightarrow 0}\,, \label{expn}
 \end{gather} 
 and since \(G\) should approach zero as \(\la^{(1-\alpha)d/\Delta_{\la}}\) we find that necessarily 
 \begin{gather}
 	j_{1}\cdots j_{n}\partial_{y_{j_{1}}\dots y_{j_{n}}} K(y_{1},\dots,y_{d})\Big{|}_{y_{i}\rightarrow d}=0\,, \quad n\leq-(1+\al d/\Delta_{\la})\,, \quad \al d/\Delta_{\la} \in \mathds{Z}_{<0}\,, \label{gravcon}
 \end{gather}      
 where we used that for \(n\geq 1\), we have \(\tensor{f}{^{\m}_{a}}\dd\tensor{e}{_{\mu}^{a}}/d\la\neq 0\) generally when \(\la\rightarrow 0\). From there we understand that only the \(n=0\) condition is truly universal, since it holds independently of the inequation we just mentioned.  Notice that when \(\al d/\Delta_{\la}\)  is not a negative integer, the term proportional to \(\la^{(1-\alpha)d/\Delta_{\la}}\) is unreachable via the standard Taylor expansion, thus one needs to employ other methods.\footnote{One can imagine that instead of matching the exact powers of \(\la\) when the negative integer condition for \( \al d/\Delta_{\la} \) is not satisfied, we push to the closest integer value such that \(G\) approaches zero slightly faster than \(\la^{(1-\alpha)d/\Delta_{\la}}\). While this is a perfectly valid approach to this issue, we will proceed with a simpler choice.} However, we have to stress that the whole \(T\overline{T}\) program relies on the assumption of a well-defined Taylor expansion around \(\lambda\rightarrow 0\), which in turn means that \eqref{gravcon} is indeed the simplest set of constraints on \(K\).
 
 The above covered one part of the story, let us now see what we can do for \eqref{intout}, which can be written as follows
\begin{gather}
	\bar{e}\tensor{(\bar{T}_{0})}{^{\mu}_{a}}\tensor{\bar{e}}{_{\nu}^{a}}=-f\la^{d/\Delta_{\la}}\sum_{j=1}^{d}j\partial_{y_{j}}K(y_{1},\dots,y_{d})\Big{|}_{\tensor{e}{_{\mu}^{a}}\rightarrow\tensor{\bar{e}}{_{\mu}^{a}}}\tensor{\qty(\bar{Y}_{j})}{^{\m}_{\n}}\,.  \label{eqmot} 
\end{gather}
Expanding both sides around \(\la\rightarrow 0\) and equating terms with the same power one can recursively solve for the auxiliary vielbein, but due to the explicit \(\lambda^{d/\Delta_{\la}}\) dependence we also obtain the following set of strikingly similar  constraints 
\begin{gather}
	j_{1}\cdots j_{n+1} \partial_{y_{j_{1}}\dots y_{j_{n+1}}}K(y_{1},\dots,y_{d})\Big{|}_{y_{i}\rightarrow d}=0\,, \quad n\leq-(1+d/\Delta_{\la})\,, \quad d/\Delta_{\la} \in \mathds{Z}_{<0}\,, \label{gravcon2}
\end{gather}
where we once more postulated that \(\tensor{f}{^{\m}_{a}}\dd\tensor{e}{_{\mu}^{a}}/d\la\neq 0\) as \(\la\rightarrow 0\) for \(n\geq1\). In the same manner as \eqref{gravcon}, the \(n=0\) condition is valid without this assumption and \(d/\Delta_{\la}\) must be a negative integer since either side of \eqref{eqmot} is expected to have a smooth polyonymic expansion. Actually, the \(n=0\) for \eqref{gravcon2} shows that \eqref{gravcon} is valid in general and without any assumptions for \(n=0\) and \(n=1\).

We are now ready to put everything together and completely determine how the initial conditions should behave based on our previous analysis. If we truly want to work with a universal gravitational dressing, its initial conditions cannot depend on the choice of \(S_{0}\), which affects \(\tensor{e}{_{\mu}^{a}}\) via \eqref{intout}, which then appears in the expansion \eqref{expn} and from there in the resulting constraints. This fact reduces the allowed values  of \(n\) to only \(n=0,1\) for \eqref{gravcon} and the value \(n=0\) for \eqref{gravcon2}. But these choices uniquely determine \(\al\) and \(\Delta_{\la}\) which now become
\begin{gather}
	\al=2\,, \qquad \Delta_{\la}=-d\,. \label{choi}
\end{gather}    
Using these, we finally arrive at
\begin{gather}
	K(d,\dots,d)=0\,,\qquad  j\partial_{y_{j}}K(y_{1},\dots,y_{d})\Big{|}_{y_{i}\rightarrow d}=0\,, \label{finin}
\end{gather} 
where the first equation is essentially \eqref{inco} and the second one is a direct result of the assumption of a well-behaved expansion of \eqref{intout}. It is worth noting that the requirements above essentially force the expansion of \(G\) to initiate at linear order in \(\la\), therefore if one would be interested in an expansion that begins at \(\la^{n}\) where \(n \in \mathds{Z}_{>0}\), the initial conditions generalize to
\begin{gather}
	K(d,\dots,d)=0\,,\qquad 	j_{1}\cdots j_{n}\partial_{y_{j_{1}}\dots y_{j_{n}}} K(y_{1},\dots,y_{d})\Big{|}_{y_{i}\rightarrow d}=0\,, \qquad n\geq1\,,
\end{gather}
when \eqref{choi} is valid.

These results will be sufficient for us to proceed to the next sections, where we will confirm their validity and provide new insights about stress-tensor deformations in various dimensions.
\section{Implications in two dimensions} \label{sec3}
The main aim of this section is to solve the flow equation \eqref{floweq}. To achieve this, we begin by specifying the generic operator that will be used for the entirety of this section
\begin{gather}
	\mathcal{O}=\int\dd[2]{x}f\hspace{0.05cm}\qty{b_{0}\tensor{T}{^\mu_\nu}\tensor{T}{^\nu_\mu}+b_{1}(\tensor{T}{^\mu_\mu})^{2}+\frac{b_{2}}{\la}\tensor{T}{^\mu_\mu}+\frac{b_{3}}{\la^{2}}}\,, \label{2ddef}
\end{gather}
with \((b_{0},b_{1},b_{2},b_{3})\) being dimensionless coefficients due to \eqref{choi}. Considering \eqref{gdef}, the corresponding ansatz for the massive gravity potential will be 
\begin{gather} \label{quad ansatz}
	S_{G}=\frac{1}{\la}\int\dd[2]{x}f\,\{c_{0}+c_{1}y_{1}+c_{2}y_{1}^{2}+c_{3}y_{2}\}\,.
\end{gather}
Using only \eqref{pde} and without specifying any initial conditions yet, we reach the following solution\footnote{Here \(b_{0}\neq0\). The case where \(b_{0}=0\) will be studied in subsection \ref{sectr}.}
\begin{gather}\label{eq:sols in 2d}
	\begin{gathered}
		b_{1}=-b_{0}\,, \qquad b_{3}=-\frac{b_{2}(1+b_{2})}{2b_{0}}\,, \\
		c_{0}=\frac{1+b_{2}}{2b_{0}}\,,\qquad c_{2}=b_{0}c_{1}^{2}\,,\qquad c_{3}=-c_{2}\,,
	\end{gathered}
\end{gather}
where we notice that \((b_{0},b_{2})\) and \(c_{1}\) are still undetermined. After a closer inspection, we realize that \(b_{0}\) can be reabsorbed by \(\la\) since it represents the scaling freedom of the deformation parameter. Hence, without any loss of generality we fix it to \(b_{0}=1/2\) for convenience, leaving us with \(b_{2}\) and \(c_{1}\) which will be determined uniquely by \eqref{finin}. Using the Neumann boundary condition, we recover \(c_{1}=-1\) so the operator and gravity action assume the following form (in matrix notation)
\begin{gather}
	\mathcal{O}=\int\dd[2]{x}f\hspace{0.05cm}\qty{-\det T+\frac{b_{2}}{\la}\tr T-\frac{b_{2}(1+b_{2})}{\la^{2}}}\,, \quad S_{G}=\frac{1}{\la}\int\dd[2]{x}f\,\qty{1+b_{2}-y_{1}+\det Y_{1}}\,. \label{opp2d}
\end{gather}
Finally, the Dirichlet condition of \eqref{finin} fixes the remaining constant to be \(b_{2}=0\), identifying \(T\overline{T}\) as the operator and \(S_{G}\) as ghost-free massive gravity \cite{Freidel:2008sh,Tolley:2019nmm,Mazenc:2019cfg}. It follows that the deformed partition function for a flat background is then given by \cite{Dubovsky:2017cnj,Dubovsky:2018bmo}
\begin{gather}
	\mathcal{Z}_{\la}=\int\frac{\mathcall{D}e\mathcall{D}\mathcal{Y}\hspace{0.02cm}}{\text{vol}(\text{diff})}\exp{\frac{i}{2\la}\int\dd[2]{x}\tensor{\e}{^{\m}^{\n}}\tensor{\e}{_{a}_{b}}(\tensor{e}{_{\m}^{a}}-\partial_{\mu}\mathcal{Y}^{a})(\tensor{e}{_{\n}^{b}}-\partial_{\nu}\mathcal{Y}^{b})}\mathcal{Z}_{0}[\tensor{e}{_{\m}^{a}}]\,, \label{path1}
\end{gather}
where we restored diffeomorphism invariance by introducing the diffeomorphism Stückelberg fields \(\mathcal{Y}^{a}\), \(\text{vol}(\text{diff})\) is the volume of the reparametrization group and \(\mathcal{Z}_{0}\) the undeformed partition function.

 We must note that the derived operator above with arbitrary \(b_{2}\) has a string theoretic interpretation once we derive its spectrum \cite{Tsolakidis:2024wut}, by considering the deformation of a CFT on a cylinder of circumference \(R\). Briefly, one finds
\begin{gather}
	E^{\pm}_{n}=-\frac{R\qty(1+2b_{2})}{2\la}\pm\sqrt{\qty(\frac{2\pi p_{n}}{R})^{2}+\qty(
		\frac{R}{2\la})^{2}+\frac{\mathcal{E}_{n}}{\la}}\,, \label{speccc}
\end{gather}
where \(p_{n}\in\mathds{Z}\) and \(\mathcal{E}_{n}/R\) corresponds to the undeformed energy levels. Proceeding with the following identifications
\begin{gather}
	\la\leftrightarrow\dfrac{\al'\pi}{w}\,, \quad b_{2}\leftrightarrow\dfrac{\tensor{B}{_{0}_{1}}-1}{2}\,, \quad p_{n}\leftrightarrow\dfrac{N_{\text{L}}-N_{\text{R}}}{w}\,, \quad \mathcal{E}_{n}\leftrightarrow\dfrac{2\pi\qty(N_\text{L}+N_{\text{R}}+k_{\bot}^{2})}{w}\,, \label{identific}
\end{gather}
the spectrum above matches the one of a closed string in a \(B\)-field \cite{Danielsson:2000gi} when \(p_{\bot}^{2}=2k_{\bot}^{2}/\al'\). The role of \(b_{2}\) is now clear, it corresponds to a non-trivial value for the \(B\)-field. With this in mind, we move on to the next subsection where we will verify independently via perturbation theory that the operator \eqref{opp2d} necessarily assumes this form.    
\subsection{Perturbation theory for \(T\overline{T}+ \tr T +\Lambda_{2}\)}
Here, we will show that one does not need to write down the specific ansatz of \eqref{sol1} for the deformed action, i.e. using the massive gravity paradigm, in order to independently reproduce the operator \eqref{opp2d}. One can instead write down the most generic deformed action of a two-dimensional theory in terms of the Lagrangian density
\begin{gather}\label{def action}
	S_{\la}=\int\dd[2]{x} f\hspace{0.05cm}\La_{\la}\,,\qquad 	\La_{\la}=\sum_{n=-1}^{\infty}\la^{n} \La_{n}\,, 
\end{gather}
where we wrote the Lagrangian density as a Laurent expansion\footnote{Please note that the operator \eqref{2ddef} implies that terms proportional to \(\la^{-1}\) will also be part of the expansion.} in the deformation parameter $\la$ and the only assumption we make is that the Lagrangian density of the undeformed theory is given by $\La_{0}$.
In this language the stress-tensor is given by
\begin{gather}\label{def stress tensor}
	\tensor{T}{^{\mu}_{\nu}}=\sum_{n=-1}^{\infty}\la^{n} \tensor{(T_{n})}{^{\mu}_{\nu}}\,,\qquad \tensor{(T_{n})}{^{\mu}_{\nu}}\coloneqq\La_{n}\tensor{\delta}{^{\mu}_{\nu}}+\pdv{\La_{n}}{\tensor{f}{_{\mu}^{a}}}\tensor{f}{_{\nu}^{a}}\,,
\end{gather}
and in matrix notation its determinant reads
\begin{gather}\label{def det stress}
	\det T=\sum_{n,m=-1}^{\infty}\la^{n+m}\det T_{n,m}\,,\qquad  \det T_{n,m}\coloneqq\frac{1}{2}\qty(\tr T_{n}\tr T_{m}-\tr(T_{n}T_{m}))\,.
\end{gather}
Moreover, the relation we obtain from the flow equation \eqref{floweq} when the deforming operator is \eqref{2ddef} with $b_{0}=-b_{1}=1/2$ and $b_{2}, b_{3}$ undetermined is 
\begin{gather}\label{eq:flow pert}
	\sum_{n=-1}^{\infty}n\la^{n-1}\La_{n}=-	\sum_{n,m=-1}^{\infty}\la^{n+m}\det T_{n,m}+b_{2}\sum_{n=-1}^{\infty}\la^{n-1}\tr T_{n}+b_{3}\la^{-2}\,.
\end{gather}
We continue by solving this equation order by order in powers of $\la$. At $\la^{-2}$ we arrive at a PDE for $\La_{-1}$ due to the relation of $T_{-1}$. However, its solution requires further initial/boundary conditions which we do not have as we only know about \(\La_{0}\). For this reason, the only acceptable solution for \(\La_{-1}\) is the one that does not implicitly depend on the background \(\tensor{f}{_{\mu}^{a}}\). Then, the PDE reduces to a simple algebraic equation which can be written as
\begin{gather}
	b_{3}=\La_{-1}(\La_{-1}-2b_{2}-1)\,.
\end{gather} 
On the other hand, at $\la^{-1}$ one can easily show that $b_{2}=\La_{-1}$ for the same reasons and thus	$b_{3}=-b_{2}(1+b_{2})$, matching the operator in \eqref{opp2d} on the nose! Having determined the negative powers in \(\la\), the remaining equations  \eqref{eq:flow pert} at positive powers in $\la$ reduce to the known perturbative solution for \(T\overline{T}\) \cite{Cavaglia:2016oda}
\begin{gather}
	\La_{n}=-\frac{1}{n}\sum_{m=0}^{n-1}\det T_{m,n-m-1}\,,\qquad n\geq 1\,,
\end{gather}
which is completely specified in terms of the undeformed Lagrangian density $\La_{0}$. 
\subsection{Hypergeometric is geometric}\label{geometric}

The goal of this subsection is to investigate the algebraic properties of special functions via stress-tensor deformations of a two-dimensional QFT. To do so, we will apply the massive gravity method to two specific seed actions. The first one is given by a non-linear sigma model in a potential, while the second is an interacting Yang-Mills-like theory with Dirac spinors.  We then show how the resulting deformed action can be recast as an infinite series of hypergeometric functions and provide an infinite extension to Schwarz's list \cite{Schwarz1873}.

Let us start our analysis with the seed action of the non-linear sigma model that describes the interaction of bosons and fermions with a non-trivial potential, that is 
\begin{gather}\label{seed NLSM}
	S_{0}=-\int\dd[2]{x}e\qty{\frac{1}{2}\tensor{g}{^\mu^\nu}\tensor{\Phi}{_\mu_\nu}+\frac{1}{2}\tensor{e}{^{\mu}_{a}}\tensor{\Psi}{^{a}_{\mu}}+V}\,, 
\end{gather}
where the bosonic and fermionic kinetic terms are respectively contained in\footnote{Here we define the symmetrization of the indices as \(\tensor{A}{_{(\mu}_{\nu)}}\coloneqq\frac{1}{2!}(\tensor{A}{_{\mu}_{\nu}}+\tensor{A}{_{\nu}_{\mu}})\) and the antisymmetrized covariant derivative as \(A\vecev{\nabla}_{\mu}B\coloneqq A\vec{\nabla}_{\mu}B-A\cev{\nabla}_{\mu}B\).} 
\begin{gather}\label{kin terms of seed action}
	\tensor{\Phi}{_\mu_\nu}\coloneqq \tensor{A}{_i_j}\partial_{(\mu}\phi^{i}\partial_{\nu)}\phi^{j}\,,\qquad	\tensor{\Psi}{^{a}_{\mu}}\coloneqq \tensor{B}{_i_j}\bar\psi^{i}\gamma^{a}\vecev{\nabla}_{\mu}\psi^{j}=\tensor{B}{_i_j}\bar\psi^{i}\gamma^{a}\vecev{\partial}_{\mu}\psi^{j}\,,
\end{gather}
with $\tensor{A}{_i_j}/\tensor{B}{_i_j}$ being arbitrary matrices. More generally, topological terms can be ignored at this stage.
Regarding the covariant derivative that appears in \eqref{kin terms of seed action}, we have
\begin{gather} \label{cov deriv}
	\nabla_{\mu}\psi^{i}=(\partial_{\mu}+\frac{1}{4}\tensor{\omega}{_{\mu}^a^b}\tensor{\gamma}{_a_b})\psi^{i}\,,\qquad
	\psi^{i}\cev\nabla_{\mu}=\psi^{i}(\cev \partial_{\mu}-\frac{1}{4}\tensor{\omega}{_{\mu}^a^b}\tensor{\gamma}{_a_b})\,,
\end{gather}
where $\tensor{\omega}{_{\mu}^a^b}$ is the spin connection. 
Previously, the \(T\overline{T}\) deformation of \eqref{seed NLSM} and certain limits thereof were studied using different methods in \cite{Cavaglia:2016oda,Bonelli:2018kik,Kraus:2018xrn,Frolov:2019nrr,Coleman:2019dvf,Caputa:2020lpa}. For the dressing \eqref{opp2d} with \(b_{2}=0\) the solution of \eqref{intout} for the auxiliary zweibein reads
\begin{gather}\label{zweibein sol NLSM}
	\tensor{(e_{\pm})}{_{\mu}^{a}}=\frac{1}{2(1-V\la)}\qty(\tensor{\hat{f}}{_{\mu}^{a}}\pm\frac{\tensor{\hat{f}}{_{\mu}^{a}}+2\la (1-V\la)\tensor{\Phi}{_{\mu}_{\nu}}\tensor{\hat{f}}{^{\nu}_{b}}\tensor{\eta}{^b^a}}{\sqrt{\widehat\gamma^{-1}\det(\tensor{\widehat\gamma}{_\mu_\nu}+2\la(1-V\la)\tensor{\Phi}{_\mu_\nu})}})\,,
\end{gather}
where we defined
\begin{gather}
	\tensor{\hat f}{_{\mu}^{a}}\coloneqq\tensor{f}{_{\mu}^{a}}+\frac{\la}{2}\tensor{\Psi}{_{\mu}^{a}}\,,\qquad \widehat{\gamma}_{\mu\nu}\coloneqq \tensor{\hat f}{_{\mu}^{a}}\tensor{\eta}{_a_b}\tensor{\hat{f}}{_{\nu}^{b}}\,.
\end{gather}
Let us now restrict to the case of a single boson without fermions.  We choose to study the deformed action one obtains by considering the positive branch of the solution \eqref{zweibein sol NLSM} as this is the one with a smooth \(\la\rightarrow 0\) limit. After some simplifications we get
\begin{gather}\label{def action NLSM}
	S_{\la}=\int\dd[2]{x}f\La_{\la}=\int\dd[2]{x}f\Bigg\{\frac{1}{\la}-\frac{1}{2\la(1-V\la)}\qty(1+ \sqrt{1+2\la(1-V\la)X})\Bigg\}\,,
\end{gather}
where we introduced the shorthand notation $X\coloneqq\tensor{\gamma}{^\mu^\nu}\tensor{\Phi}{_\mu_\nu}$.  

We proceed by obtaining the deformed Lagrangian density via a perturbative expansion in $\la$. 
However, it should not matter whether one studies the expansion around $\la\rightarrow0$ or instead expands for small $V$. The reason this latter expansion can be performed without loss of generality is because the potential always appears in the deformed action as a product with the deformation parameter $\la$, i.e. in the form of $V\la$. By considering the aforementioned approach, we find the following answer\footnote{This expression is equivalent to the one found in \cite{Cavaglia:2016oda}.} for the deformed Lagrangian density
\begin{gather}
	\La_{\la}=-\frac{V}{1-V\la}-\frac{X}{2}\sum_{n=0}^{\infty}(2n-1)!!(\la^{2}V X)^{n}\tpFq{2}{1}{\frac{1}{2}+n,1+n}{2+n}{-2\la X}\,,
\end{gather}
where we defined the tilded hypergeometric function as the regularized hypergeometric function via
\begin{gather}\label{hyper reg}
	\tpFq{p}{q}{a_{1},a_{2},\dots,a_{p}}{b_{1},b_{2},\dots,b_{q}}{c}\coloneqq\frac{1}{\Gamma(b_{1})\Gamma(b_{2})\cdots\Gamma(b_{q})}\pFq{p}{q}{a_{1},a_{2},\dots,a_{p}}{b_{1},b_{2},\dots,b_{q}}{c}\,.
\end{gather}
By appropriately combining the above results we find that the infinite sum of hypergeometric functions resums to the following expression
\begin{gather}\label{sum hypers from NLSM}
	\sum_{n=0}^{\infty}(2n-1)!!(yw)^{n}\tpFq{2}{1}{\frac{1}{2}+n,1+n}{2+n}{-2w}=	\frac{\sqrt{1+2w(1-y)}-1}{w(1-y)}\,,
\end{gather}
where we introduced the variables $y\coloneqq V\la$ and $w\coloneqq X\la$. Expanding now the right hand side for small \(y\), one can obtain the algebraic form for this particular \({}_{2}\widetilde{F}_{1}\) family by comparing both sides of \eqref{sum hypers from NLSM} for every \(n\).  In this case, one retrieves already known results.

In our above analysis we restricted to the case of a single boson. However, it would be interesting to extend our results to the case of $N$ bosons and examine whether this could yield further insight into the closed form expression in terms of radicals, of particular infinite sums of special functions.  We leave this for future studies.

Let us now switch gears and continue our analysis by studying a different seed action, namely 
\begin{gather}\label{seed F+V+Psi}
	S_{0}=-\int\dd[2]{x}e\qty{\frac{1}{4}\text{tr}(\tensor{F}{_\mu_\nu}\tensor{g}{^\mu^\rho}\tensor{F}{_\rho_\lambda}\tensor{g}{^\la^\nu})+\frac{1}{2}\tensor{e}{^{\mu}_{a}}\tensor{\Psi}{^{a}_{\mu}}+V}\,,
\end{gather}
where the trace is taken over the suppressed gauge group indices. Here, the covariant derivative $\tensor{\nabla}{_\mu}$ of \eqref{cov deriv} now also contains a connection term coming from the gauge field.
Finally, assuming that the field strength is the usual antisymmetric two-tensor, the subsequent analysis holds for both abelian and non-abelian gauge fields. Considering the factorization of the deformed action \eqref{sol1}, this argument is also valid for any other stress-tensor deformation.

With the above in mind, we may now once again solve the equations of motion coming from \eqref{opp2d} with $b_{2}=0$, together with the above seed action to extract a solution for the auxiliary zweibein; that is 
\begin{gather}\label{esol F V Psi}
	\tensor{(e_{s,\pm})}{_{\mu}^{a}}=\frac{H_{s,\pm}}{1-\la V}\tensor{\hat f}{_{\mu}^{a}}\,,\qquad 	\tensor{\hat f}{_{\mu}^{a}}\coloneqq\tensor{f}{_{\mu}^{a}}+\frac{\la}{2}\tensor{\Psi}{_{\mu}^{a}}\,,
\end{gather}
with 
\begin{gather}\label{Hsols}
	H_{s,\pm}=\frac{1}{4}\Bigg(1+s\sqrt{\pm\frac{2}{\sqrt{x}}+3-x}\pm\sqrt{x}\Bigg)\,,\qquad s=\pm1\,,
\end{gather}
where we defined 
\begin{gather}\label{x, P and Fsq defs}
	\begin{gathered}
		x\coloneqq\frac{3P}{2(P+\sqrt{P^{2}-P^{3}})^{1/3}}+\frac{3}{2}(P+\sqrt{P^{2}-P^{3}})^{1/3}+1\,,\\
		P\coloneqq\frac{64F^{2}\la}{27(f^{-1}\hat f)^{2}}(1-V \la)^{3}\,, \qquad F^{2}\coloneqq\text{tr}(\tensor{F}{_\mu_\nu}\tensor{\gamma}{^\mu^\rho}\tensor{F}{_\rho_\lambda}\tensor{\gamma}{^\la^\nu})\,.
	\end{gathered}
\end{gather}
Moreover, the requirement of the auxiliary metric to be positive definite implies that $H\in \mathds{R}$. This is true for $H_{s,+}$ and yields $0< x\leq 4$, $-\infty < P\leq 1$, while the two remaining solutions $H_{s,-}$ are purely imaginary or zero. These ranges imply that $F^{2}$ must be bounded by a critical value, that is
\begin{gather}\label{Fsq crit}
	\frac{64 F^{2} \la (1-V \la)^{3}}{27(f^{-1}\hat f)^{2}}\leq 1\,.
\end{gather}
Notably, in the $V\rightarrow0$ limit and for the case where the fermions are turned off, the upper bound of \eqref{Fsq crit} agrees with the critical electric field that was found in \cite{Brennan:2019azg}. 
Inserting the above solution into the deformed action \eqref{sol1} gives
\begin{gather}\label{def action F+V+Psi}
	S_{\la}=\int\dd[2]{x}f\Bigg\{\frac{1}{\la}-\frac{F^{2}(1-V\la)^{2}}{4f^{-1}\hat f H_{s,\pm}^{2}}+\frac{f^{-1}\hat fH_{s,\pm}(H_{s,\pm}-2)}{\la(1-V\la)}\Bigg\}\,.
\end{gather}
  Clearly, when we turn off the deformation parameter in \eqref{esol F V Psi} we must recover the initial condition given by \eqref{inco}. This is achieved when $H\lvert_{\la\rightarrow 0}= 1$, which is only true for the $H_{1,+}$ solution and hence is the one we should analyse.

Similarly as before, we can continue our analysis by turning off the fermions. Certainly, upon inserting the appropriate solution of \eqref{Hsols}, this action becomes highly non-trivial as it is given in terms of (powers of) cubic roots. In addition, this answer is not very intuitive. In an effort to better understand this solution we utilize the perturbative expansion outlined in the previous subsection. For our purposes, however, this is not enough and one needs to perform an additional expansion around $V\rightarrow0$. As explained above, this can be done without any loss of generality, since once again the action depends on the potential only through the product $V\la$. This analysis allows us to resum the full perturbative expansion in $\la$ to a hypergeometric function. In turn, the perturbative expansion in the potential is then given by an infinite sum of hypergeometric functions.  

More concretely, we find that the deformed Lagrangian of \eqref{def action F+V+Psi} without fermions can be written as follows
\begin{gather}\label{def L YM}
	\mathcal{L}_{\la}=-\frac{V}{1-V\la}-\pi F^{2}\sum_{n=0}^{\infty}\frac{3^{n-5/2}(-V\la)^{n}}{\Gamma(1+n)}\tpFq{5}{4}{\frac{1}{2},\,\,\frac{3}{4},\,\,1,\,\,1,\,\,\frac{5}{4}}{2,\frac{3-n}{3},\frac{4-n}{3},\frac{5-n}{3}}{\tfrac{64}{27}F^{2}\la}\,,
\end{gather}
where the regularized hypergeometric function is related to the usual hypergeometric function through \eqref{hyper reg}.  Notice that in the limit of $V\rightarrow 0$ we recover the deformed action that was obtained in \cite{Brennan:2019azg}. 
By comparing \eqref{def L YM} to the closed form expression of \eqref{def action F+V+Psi} with no fermions, we find that the infinite sum of (regularized) hypergeometric functions can be written in terms of radicals, viz.
\begin{gather}\label{hyper 2}
	\sum_{n=0}^{\infty}\frac{3^{n-5/2}(-y)^{n}}{\Gamma(1+n)}\tpFq{5}{4}{\frac{1}{2},\,\,\frac{3}{4},\,\,1,\,\,1,\,\,\frac{5}{4}}{2,\frac{3-n}{3},\frac{4-n}{3},\frac{5-n}{3}}{\tfrac{64}{27}z}=\frac{(1-y)^{2}}{4\pi \mathcal{H}^{2}}-\frac{(\mathcal{H}-1)^{2}}{\pi z(1-y)}\,,
\end{gather}
where we defined $\mathcal{H}\coloneqq H_{1,+}$ with the fermions turned off, $y\coloneqq V\la$ and $z\coloneqq F^{2}\la$, to clarify that this results in a Taylor expansion around $y=0$ on the left hand side of the equation. 

Remarkably, combining this expression, with the previous result \eqref{sum hypers from NLSM} implies that stress-tensor deformations can in general be used as a tool for deriving closed form expressions of infinite sums of special functions when using the massive gravity approach. From this perspective hypergeometric functions are geometric.

Equation \eqref{hyper 2} can provide yet another infinite extension of Schwarz's list as one can Taylor expand in $y$ the right hand side of each equation and compare both sides order by order. Then we can directly extract the algebraic form of the corresponding hypergeometric function. It is important to stress once more that the massive gravity approach is the reason as to why these special functions can be recast in terms of these algebraic expressions.

Before we proceed to the next subsection a few comments are in order. First, throughout our latter analysis for the $T\overline{T}$ deformation of \eqref{seed F+V+Psi}, we considered the case without fermions but with a non-trivial potential. Although this choice is not very physical, it generated a closed form expression for an infinite sum of hypergeometric functions. Hence, we argue that even the study of seed actions that (perhaps) lack physical significance, could lead to fresh insights regarding the study of such infinite sums. Nonetheless, we aim to include fermions in a subsequent analysis. Second, it would be interesting to combine the two previous seed actions into one action functional, such as
\begin{gather}\label{def S YM+NLSM}
	S_{0}=-\int\dd[2]{x}e\qty{\frac{1}{4}\text{tr}(\tensor{F}{_\mu_\nu}\tensor{g}{^\mu^\rho}\tensor{F}{_\rho_\lambda}\tensor{g}{^\la^\nu})+\frac{1}{2}\tensor{g}{^\mu^\nu}\tensor{\Phi}{_\mu_\nu}+\frac{1}{2}\tensor{e}{^{\mu}_{a}}\tensor{\Psi}{^{a}_{\mu}}+V}\,,
\end{gather}
and examine whether performing the above analysis would generate an algebraic solution for other hypergeometric functions. Looking at \eqref{def S YM+NLSM} and restricting to the case of a single boson with no fermions yields equations of motion for the zweibein that assume the form of a sixth degree polynomial. This means that a solution in radicals does not exist. Therefore, the resulting infinite sum, which may again contain special functions, will not have an algebraic description. This fact shows that this approach can be thought of as both a go or no-go theorem!

The algorithm we followed to express certain hypergeometric functions and their infinite sums in terms of algebraic functions can be summarized as follows. First, we write down a seed action and combine this with the massive gravity action \eqref{opp2d}. From there we can extract the equations of motion for the auxiliary zweibein. If the resulting equations are a polynomial of degree less than five, then the zweibein admits a solution in radicals. Otherwise, this will not be the case. In the former, it then proves useful to perform a perturbative analysis of the corresponding flow equation, we expect that this allows one to express certain special functions algebraically.

On the other hand, starting from a special function, say $\Sigma(u)$, it is interesting to construct a reverse algorithm. This will allow one to express $\Sigma(u)$ in terms of an algebraic function. Essentially, this boils down to finding the correct scalar quantity $u$ to identify with the appropriate combination of (background) fields. In our previous examples this scalar was given by $w, z$. More generally, however, this scalar can be of the following form 
\begin{gather}
	u\coloneqq \la \La_{0}= \la\, (\text{background})\cdot(\text{fields})=\la\,\mathfrak{f}\cdot \mathfrak{u}\,,
\end{gather}
where we suppressed the index structure of $\mathfrak{f}$ and $\mathfrak{u}$, each of which can include appropriate combinations of the vielbein, spin connection,  gauge field, bosonic fields, fermionic fields and possible derivatives thereof.\footnote{Recall that in the massive gravity approach $\mathfrak{f}$ only contained appropriate combinations of the vielbeine.} In this language, the deformed action and operator read
\begin{gather}
	S_{\la}[\bar{\mathfrak{e}}, \mathfrak{f},  \la, \mathfrak{u}]=\int\dd[d]{x}f\, \Sigma(u)\,,\qquad	\mathcal{O}=\int\dd[d]{x}f\,F(\mathfrak{J})\,, \qquad \mathfrak{J}\coloneqq\frac{1}{f}\frac{\de S_{\la}}{\de \mathfrak{f}}\,,
\end{gather}
where $\mathfrak{e}$ is the auxiliary background field and $\bar{\mathfrak{e}}$ denotes its on-shell value, we will come back to this momentarily. 
Before proceeding, we highlight that one must find an appropriate $F(\mathfrak{J})$, such that \eqref{floweq} holds, i.e. $\partial_{\la}\Sigma(u)=F(\mathfrak{J})$ which, in principle, is not an easy task.

Assuming that this can be done,  we introduce the following factorization of the deformed action
\begin{gather} \label{def S gen}
	S_{\la}[\mathfrak{e}, \mathfrak{f},\la, \mathfrak{u}]=S_{G}[\mathfrak{e}, \mathfrak{f}, \la]+S_{0}[\mathfrak{e},\mathfrak{u}]=\int\dd[d]{x}f\,G(\la,\mathfrak{e}, \mathfrak{f})+\int\dd[d]{x}e\,\La_{0}(\mathfrak{e},\mathfrak{u})\,,
\end{gather}
and find a solution for $\partial_{\la}G(\la,\mathfrak{e}, \mathfrak{f})=F(\mathfrak{J})$, that satisfies the appropriate initial conditions when $\la\rightarrow0$, in analogy with \eqref{finin}. From here it follows that one must integrate out the auxiliary field, as in \eqref{condtt}. As mentioned earlier, if the resulting equations that one recovers for $\mathfrak{e}$, upon solving the equations of motion, are polynomials of degree less than five, one can obtain a solution for $\bar{\mathfrak{e}}=\bar{\mathfrak{e}}(\la,\mathfrak{f},\mathfrak{u})$ in radicals. Evaluating the deformed action on-shell will in turn yield a Lagrangian density that can be written as an algebraic function. Following these steps could then lead to the algebraic formulation of a given special function $\Sigma(u)$. Furthermore, our examples indicate that the introduction of an arbitrary potential $V$ in the seed action will allow one to obtain such expressions for infinite sums of special functions, when these take the form of a Taylor expansion around small $V$.

\subsection{The local renormalization group and the trace-flow equation} \label{sec3.3}
The trace-flow equation relates the expectation value of the \(T\overline{T}\) operator to the expectation value of the trace of the deformed stress-tensor. This was initially noted in \cite{Cavaglia:2016oda} at the classical level and in \cite{McGough:2016lol} quantum mechanically. For this subsection, we will explain how this result is essentially a special case of a known local RG flow equation \cite{Drummond:1977dg,Shore:1986hk,Osborn:1987au,Osborn:1988wu,Osborn:1989td,Jack:1990eb,Osborn:1991gm,Peskin:1995ev} (see \cite{Shore:2016xor} for a nice review) which will allow us to compute the corresponding beta function up to all orders.

First, let us assume that our theory contains some couplings along with their scaling dimensions \((\lambda_{i}, \Delta_{\la_{i}})\), some fields \((\Phi_{i},\Delta_{\Phi_{i}})\) and the background vielbein \((\tensor{f}{_{\mu}^{a}},1)\). The next step is to temporarily uplift the couplings to sources\footnote{Under the RG flow, this will generate derivative terms that will contribute to the conformal anomaly, however in the end we will consider the limit \(\varlambda_{i}(x)\rightarrow\la_{i}\) which will make them disappear. This effect is analogous to the one we get when uplifting from a flat background to a curved one for a CFT; the price we pay is the conformal anomaly which would otherwise vanish in the flat spacetime limit.} i.e. \(\la_{i}\rightarrow\varlambda_{i}(x)\) allowing us to write down the following set of Weyl transformations (no sum)
\begin{gather}
	\tensor{f}{_{\mu}^{a}}(x)\rightarrow e^{\sigma(x)}\tensor{f}{_{\mu}^{a}}(x)\,,\qquad \Phi_{i}(x)\rightarrow e^{-\Delta_{\Phi_{i}} \sigma(x)}\Phi_{i}(x)\,,\qquad\varlambda_{i}(x)\rightarrow e^{-\Delta_{\la_{i}} \sigma(x)}\varlambda_{i}(x)\,. \label{rq1}
\end{gather}
The variation of the deformed action under \eqref{rq1} reads (yes sum)
\begin{gather}
	\delta_{\sigma} S_{\varlambda_{i}}=\int\dd[d]{x}f\qty{\frac{1}{f}\frac{\de S_{\varlambda_{i}}}{\de 	\tensor{f}{_{\mu}^{a}}}	\tensor{f}{_{\mu}^{a}}-\frac{\Delta_{\Phi_{j}}}{f}\frac{\de S_{\varlambda_{i}}}{\de\Phi_{j}}\Phi_{j}-\frac{\Delta_{\la_{j}}}{f}\frac{\de S_{\varlambda_{i}}}{\de\varlambda_{j}}\varlambda_{j}}\sigma\,,
\end{gather}
therefore from \eqref{floweq} and for a theory that is symmetric\footnote{This almost always requires the addition of curvature terms \cite{Birrell:1982ix}.} under \eqref{rq1} one finds 
\begin{gather}
	\tr T=\Delta_{\la_{i}}\la_{i}\mathcall{O}_{i} + \text{curvature/eom terms}\,,  \label{trflowclass}
\end{gather}
where \(\mathcall{O}_{i}\) are the integrand operators. The above is a classical equation, essentially showing that when the scaling dimension or couplings are absent, the theory is classically conformal on-shell\footnote{That is if \(\Delta_{\Phi_{i}}\neq0\).} for a flat background \cite{Karananas:2015ioa}. We must stress that this equation is a result of the already deformed theory. For example, consider the deformation of a CFT only via one operator with coupling \(\la\). Combining \eqref{floweq} and \eqref{trflowclass} one may be tempted to write an equation that looks like \(\Delta_{\la}\la\partial_{\la}S_{\la}=\int\dd[d]{x}f\,\tr T\) and then attempt to use it as the definition for the deformation. This will however lead to trivial results for \(S_{\la}\). Nevertheless, irrespectively of the deformation, this equation is always a good consistency check for the deformed theory. 

In order to derive the quantum version of \eqref{trflowclass}, we first observe that the Weyl rescaling of the background \eqref{rq1} takes us to larger distances (we pick \(\sigma>0\) without any loss of generality) therefore the RG flow we consider is from the UV to the IR. In more detail, if our bare theory is defined at \(\Lambda_{0}(x)\), the RG flow will take us to \(\Lambda(x)=e^{-\sigma(x)}\Lambda_{0}(x)\) and as a consequence \((\Phi_{i},\varlambda_{i})\rightarrow(\Phi^{\text{R}}_{i},\varlambda^{\text{R}}_{i})\) i.e. the resulting effective action is now described in terms of the renormalized fields and sources. Moreover, the partition function is unaffected by this change leading to \(\mathcal{Z}_{\Lambda}[\varlambda^{\text{R}}_{i}]=\mathcal{Z}_{\Lambda_{0}}[\varlambda_{i}]\) from which one can obtain a version of the Callan-Symanzik equation. We can now study the variation of the effective action under an infinitesimal change in scale which can be expressed entirely in terms of \(\sigma\) or equivalently by an infinitesimal change of the background vielbein. Under this  transformation the renormalized fields will pick up an anomalous dimension \(\widetilde{\vargamma}_{\Phi_{i}}(x)\) and the corresponding effect for the sources is given by the beta functions \(\widetilde{\beta}_{\varlambda_{i}}(x)\), where tilde indicates the local dependence. Gathering everything we have that 
\begin{gather}
	\delta_{\sigma} S^{\text{eff}}_{\varlambda_{i}}=\int\dd[d]{x}f\qty{-\frac{1}{f}\frac{\de S^{\text{eff}}_{\varlambda_{i}}}{\de 	\tensor{f}{_{\mu}^{a}}}	\tensor{f}{_{\mu}^{a}}-\frac{\widetilde{\vargamma}_{\Phi_{j}}}{f}\frac{\de S^{\text{eff}}_{\varlambda_{i}}}{\de\Phi^{\text{R}}_{j}}\Phi^{\text{R}}_{j}+\frac{\widetilde{\beta}_{\varlambda_{j}}}{f}\frac{\de S^{\text{eff}}_{\varlambda_{i}}}{\de\varlambda^{\text{R}}_{j}}}\sigma\,, 
\end{gather}
with \(\widetilde{\vargamma}_{\Phi_{i}}\coloneqq-\partial\log\Phi^{\text{R}}_{i}/\partial\sigma\) and \(\widetilde{\beta}_{\varlambda_{i}}\coloneqq\partial\varlambda^{\text{R}}_{i}/\partial\sigma\). From there, we can finally write down the local RG flow equation 
\begin{gather}
	\tr T=\beta_{\lambda_{i}}\mathcall{O}_{i} +\text{anomalies}+ \text{curvature/contact terms}\,, \label{trflowclassquant}
\end{gather}
which can now be treated as an operator equality.\footnote{Please note that in both \eqref{trflowclass} and \eqref{trflowclassquant}, terms proportional to total derivatives can appear, however they will not affect our final result.} In that light, we now turn off all couplings but one \((\lambda)\), we pick a flat background, fix the dimension to be \(d=2\) and the choice for an operator reads
\begin{gather}
	\mathcall{O}=-2b_{0}\det T+\frac{b_{2}}{\la}\tr T+\frac{b_{3}}{\la^{2}}\,. \label{oo3rd}
\end{gather}
Under the aforementioned criteria, the seed theory is most likely a CFT, which we will pick to live on a Euclidean cylinder of circumference \(R\) with local coordinates \((\ta,x)\sim(\ta,x+R)\) and \(\ta\coloneqq it\). Similarly to \eqref{speccc}, the complete, non-perturbative quantum effect of this deformation to the undeformed spectrum is given by
\begin{gather}
	E^{\pm}_{n}=-\dfrac{R\qty(1+2b_{2})}{2\la'}\pm\sqrt{\qty(\dfrac{2\pi p_{n}}{R})^{2}+\qty(\dfrac{R\qty(1+2b_{2})}{2\la'})^{2}+\dfrac{2b_{0}\qty(b_{3}R^{2}+\la'\mathcal{E}_{n})}{{\la'}^{2}}}\,, \label{solsprec}
\end{gather}
with \(\la'\coloneqq2b_{0}\la\). Now, together with
\begin{gather}
	\ev{\tensor{T}{_\ta_\ta}}{n}=-\dfrac{E_{n}}{R}\,,\qquad 	\ev{\tensor{T}{_\ta_x}}{n}=-\dfrac{2\pi i p_{n}}{R^{2}}\,, \qquad 	\ev{\tensor{T}{_x_x}}{n}=-\pdv{E_{n}}{R}\,, \label{r2}
\end{gather}
we can evaluate the expectation value on either side of \eqref{trflowclassquant} on the basis \(\ket{n}\) and then solve for the beta function of \(\la\). Reassuringly, for both energy branches one finds a rather simple result
\begin{gather}
	\beta_{\lambda}=\Delta_{\la}\la\,, \label{betafun}
\end{gather} 
verifying that \(T\overline{T}\) is indeed a protected operator in flat space \cite{McGough:2016lol}. An equivalent way of reaching the same conclusion is to notice that from \eqref{identific}, \(\la\) can be identified with \(\al'\) which does not run under the RG flow hence the beta function would only contain the classical piece. It would be very interesting to apply these techniques to the models considered in \cite{Rosenhaus:2019utc} and carefully compare the resulting beta functions.

The main take-home message from this subsection is the subtle difference between \eqref{trflowclass} and \eqref{trflowclassquant}. The first is a classical equivalence derived by considering the generalized Weyl symmetry of the action, whereas the second is an operator equation coming from the local renormalization group procedure. A direct consequence of this discussion is that an operator satisfying \eqref{trflowclassquant} and \eqref{betafun} is either automatically protected or, a large \(N\) limit is implicitly assumed. With that in mind we can now have a clear interpretation of our previous results \cite{Tsolakidis:2024wut}  which also covers the holographic cases \cite{Taylor:2018xcy,Hartman:2018tkw}.  In short, the unique\footnote{In a sense that it corresponds to the only polynomial operator built out of finite powers of the stress-tensor that has a ghost-free massive gravity dual; more on this in subsection \ref{sec41}.} trace-flow equation in any dimension takes the following form
\begin{gather}
	\tr\bar{T}=\Delta_{\la}\la\Bigg{\{}\text{\textoverline{higher order terms}}+c_{1}\qty(\tr\bar{T}^{2}-\dfrac{1}{d-1}(\tr\bar{T})^{2})+\dfrac{c_{2}}{\la}\tr\bar{T} +\dfrac{c_{3}}{\la^{2}}\Bigg{\}}+\dfrac{\bar{e}}{f}\tr\bar{T}_{0}\,, \label{traceflowdd}
\end{gather}
where \((c_{1}, c_{2}, c_{3})\) are dimensionless constants and the last term essentially satisfies \eqref{trflowclass} for the couplings and the background of the undeformed action i.e. \(\mathfrak{g}_{i}\) and \(\tensor{e}{_{\mu}^{a}}\). This equality was derived following a strictly classical procedure which can now be understood as a manifestation of the generalized Weyl symmetry \eqref{rq1}. Then, the holographic approximation further assumes large \(N\) which automatically kills off the higher order terms. Reversing the tape, we propose (by direct comparison with \eqref{trflowclassquant}) that the full-fledged operator equality would be given by the relation above with \(\Delta_{\la_{i}}\la_{i}\rightarrow\beta_{\lambda_{i}}\) as we generally expect the couplings to run in a dimension greater than two.

As a final comment, we notice that for the case of \(T\overline{T}+\Lambda_{2}\) --where \(b_{2}=0\) and \(b_{3}\) is arbitrary in \eqref{oo3rd}-- the geometric description à la massive gravity requires the inclusion of another irrelevant coupling \(\la_{0}\) with \(\Delta_{\la_{0}}=-2\) \cite{Torroba:2022jrk}\footnote{Here we must stress that this description may not be unique, just like the quadratic ansatz in \eqref{quad ansatz}, since both are determined by a non-linear PDE.}, which would automatically generate a corresponding beta function and an operator, further complicating the trace-flow equation. To be more precise, this does not mean that such an equality does not exist for \(T\overline{T}+\Lambda_{2}\) (in fact, this is what we showed in this subsection), however if one insists on its gravitational description, extra terms will inevitably appear in the trace-flow equation, at least classically. This fact also explains why \(b_{3}\) is given by \eqref{eq:sols in 2d} under the assumption that no other couplings are present in our theory; this is the only choice allowing a massive gravity description with a trace-flow equation written entirely out of the deformed and undeformed stress-tensor.
\subsection{Initial conditions for dimensional reduction} \label{dimreduct}
In this subsection, we will verify that the gravitational description of the operator coming from the dimensional reduction of \eqref{oo3rd} matches precisely the one predicted in \cite{Gross:2019ach} only if the initial conditions \eqref{finin} are met. Interestingly, the resulting gravitational mass term will also be well-defined at the level of the partition function. 

Following the standard procedure \cite{Gross:2019ach}, the resulting operator takes the  form
\begin{gather}
	\mathcal{O}=\int\dd{x} f\, \dfrac{T^{2}+b_{3}/{\la}^{2}}{1+2b_{2}-2T\la}\,,  \label{gr1d}
\end{gather}
where we fixed \(b_{0}=1/2\). Then, from \eqref{pde} we obtain the following solution\footnote{We again assume that there are no other dimensionful couplings.}
\begin{gather}
S_{G}=\frac{1}{\la}\int\dd{x}f\,\qty{\frac{1}{2}+b_{2}+\frac{c_{0}}{y_{1}}+\frac{(1+2b_{2})^{2}+4b_{3}}{16c_{0}}y_{1}}\,. \label{gr3d}
\end{gather}
The first check is to see whether an equation like \eqref{traceflowdd} holds for the above. A direct calculation yields
\begin{gather}
	\bar{T}=\Delta_{\la}\la\dfrac{\bar{T}^{2}+b_{3}/{\la}^{2}}{1+2b_{2}-2\bar{T}\la}+\dfrac{\bar{e}}{f}\bar{T}_{0}\,, \qquad \Delta_{\la}=-1\,. \label{traceflowd11}
\end{gather}
 Applying the initial conditions \eqref{finin} we find that \(b_{3}=0\) and \(b_{2}=-(1+4c_{0})/2\), allowing us to rewrite \eqref{gr1d} and \eqref{gr3d} as
\begin{gather}
	\mathcal{O}=-\int\dd{x} f\, \dfrac{T^{2}}{4c_{0}+2T\la}\,, \qquad 	S_{G}=\frac{c_{0}}{\la}\int\dd{x}e\,(e^{-1}f-1)^{2}\,. \label{gr2d}
\end{gather}
We can now restrict to a flat background and restore diffeomorphism invariance simply by \(f=\partial_{x}\mathcal{Y}\eqqcolon\dot{\mathcal{Y}}\), where \(\mathcal{Y}\) is now the diffeomorphism Stückelberg field. The last step is to carefully compare with \cite{Gross:2019ach} achieving a perfect match for \(c_{0}=1/8\), which we understand to be just a rescaling of the deformation parameter in complete analogy with \(b_{0}\) in the two-dimensional case. Finally, the deformation at the level of the partition function reads
\begin{gather}
	\mathcal{Z}_{\la}=\int\frac{\mathcall{D}e\mathcall{D}\mathcal{Y}}{\text{vol}(\text{diff})}\exp{\frac{i}{8\la}\int\dd{x}e\,(e^{-1}\dot{\mathcal{Y}}-1)^{2}}\mathcal{Z}_{0}[e]\,, \label{path2}
\end{gather}
where its evaluation in Euclidean signature yields consistent results. The deformed energy levels simply read
\begin{gather}
	E_{n}=\frac{1-\sqrt{1-8\la\mathcal{E}_{n}}}{4\la}\,, \qquad E_{n}\Big{|}_{\la\rightarrow0}=\mathcal{E}_{n}\,,
\end{gather}
therefore with the help of \eqref{trflowclassquant} and \eqref{traceflowd11} we find that the following equation holds
\begin{gather}
	E_{n}=-\beta^{(n)}_{\la}\frac{2E_{n}^{2}}{1+4\la E_{n}}+\mathcal{E}_{n}\,, \qquad \beta^{(n)}_{\la}=\la(\sqrt{1-8\la\mathcal{E}_{n}}-2)\,, \label{betaT22}
\end{gather}
where the expansion around \(\la\rightarrow0\) gives the correct classical contribution to the beta function (assuming that this identification is valid).
Taking a step back, we notice that the described procedure of Section \ref{sec2} yields a well-behaved partition function in both one and two dimensions. 

We can now repeat the same arguments for the \(\mathcall{O}=-T^{2}\) deformation in \(d=1\), obtaining this way a gravitational dressing that one can stick into a path integral \cite{Gross:2019ach}. In this case, \eqref{pde} reduces to an ODE and applying \eqref{finin} returns a unique solution. The effect on the partition function in the diffeomorphism invariant flat space limit is now simply given by
\begin{gather}
		\mathcal{Z}_{\la}=\int\frac{\mathcall{D}e\mathcall{D}\mathcal{Y}}{\text{vol}(\text{diff})}\exp{\frac{i}{\la}\int\dd{x}e\,\qty(\sqrt{e^{-1}\dot{\mathcal{Y}}}-1)^{2}}\mathcal{Z}_{0}[e]\,, \label{path3}
\end{gather}
where we notice that a non-linear structure emerges compared to \eqref{path2}. It would be very interesting to study the properties of the kernel above, and check whether one could obtain the anticipating effect on the deformed energies. We leave this for future work. For completeness, the energy levels read
\begin{gather}
	E_{n}=\frac{\mathcal{E}_{n}}{1-\la\mathcal{E}_{n}}\,, \qquad E_{n}\Big{|}_{\la\rightarrow0}=\mathcal{E}_{n}\,.
\end{gather}
 At the classical level, we can obtain the correct trace-flow equation using \eqref{path3} i.e.
\begin{gather}
	\bar{T}=-\Delta_{\la}\la \bar{T}^{2}+\dfrac{\bar{e}}{f}\bar{T}_{0}\,, \qquad \Delta_{\la}=-1\,, \label{traceflowd12}
\end{gather}
which now suggests the following analogue for the beta functions
\begin{gather}
		E_{n}=-\beta^{(n)}_{\la}E_{n}^{2}+\mathcal{E}_{n}\,, \qquad \beta^{(n)}_{\la}=\Delta_{\la}\la+\mathcal{E}_{n}\la^{2}\,. \label{betaT222}
\end{gather} 
Another classical calculation would be to consider the effect of this deformation on say, a sigma model of bosons in a potential. Once again, this would not be a very well motivated calculation from a physical point of view, however for the case of a single boson special functions identical to the ones we retrieve in subsection \ref{geometric} make their appearance, see e.g. \cite{Gross:2019ach}. This indicates that one can use the algorithm we developed previously to study their properties and provide other types of extensions to our results.

Overall, for an operator \(\mathcall{O}(\la,T)\) the governing equation for the gravitational dressing would be given by the solutions of \eqref{pde} which reduces to an ODE when there are no other couplings present. Together with \eqref{finin}, the complete set of equations reads
\begin{gather}
K(y_{1})=-\la^{2}\mathcal{O}(\la,[K(y_{1})-y_{1}K(y_{1})']/\la)\,, \qquad 	K(1)=0\,, \qquad K(y_{1})'\Big{|}_{y_{1}\rightarrow 1}=0\,, \label{finin1d}
\end{gather}
where \(K(y_{1})'\coloneqq \dd K(y_{1})/\dd y_{1}\) and the units of \(\mathcal{O}\) are restricted to be \(M^{2}\). Building on this lower-dimensional intuition, we will repeat the same process in the next section but for higher dimensions.    
\section{Applications in higher dimensions}\label{sec4}
In this section, we will attempt to provide a direct explanation of the extra terms that appear in the higher-dimensional generalization of \(T\overline{T}\) that we propose, by understanding the physical content of the corresponding gravitational description. In turn, this will allow us to make solid and highly non-trivial predictions for the deformed field theory. We will then apply our methodology to other higher-dimensional guesses and show that they all fit very nicely into our formalism. 

\subsection{Minimal massive gravity} \label{sec41}
We will begin by recalling the structure of the unique, order \(d\) polynomial expression (in terms of \(y_{i}\)) for the gravitational dressing that corresponds to a polynomial operator of order \(d\) entirely constructed out of the deformed stress-tensor \cite{Tsolakidis:2024wut}. It is important to mention that the result was achieved after an exhaustive procedure, where the most general polynomial ansatz of order \(d\) for the massive gravity dressing and operator was used. This automatically means that we have already covered all possible cases of dRGT mass terms \cite{deRham:2010kj, deRham:2014zqa} that could have a stress-tensor deformation interpretation of the type we just mentioned. The resulting dressing is of the following form
\begin{gather}
	S_{G}=\frac{1}{\la}\int\dd[d]{x}f\,\qty{c_{0}+c_{1}y_{1}+c_{2}\det Y_{1}}\,,  \label{dressmassd}
\end{gather}
where  \((c_{0}, c_{1}, c_{2})\) are arbitrary dimensionless constants, and we will provide the corresponding operator momentarily. Applying the initial conditions \eqref{finin}, we retrieve \(c_{0}=c_{1}(1-d)\) and \(c_{2}=-c_{1}\) where \(c_{1}\) is once again an overall rescaling of \(\la\). Looking back to \eqref{opp2d}, we pick \(c_{1}=-1\) reaching
\begin{gather}
	S_{G}=\frac{1}{\la}\int\dd[d]{x}f\,\qty{d-1-y_{1}+\det Y_{1}}\,.  \label{dressmassdfixed}
\end{gather}
The first thing to notice is that for \(d=1\) this action vanishes identically, therefore one needs to consider the deformations and gravitational mass terms of subsection \ref{dimreduct}. Furthermore, since there are no other free parameters, the corresponding integrand operators are completely fixed, and their expressions are given by 
\begin{gather}
		\mathcall{O}_{d}=\frac{1}{\la^{2}}\qty[1+\la\tr\mathcal{T}-\det(\mathds{1}_{d}+\la\mathcal{T})]\,, \qquad \mathcal{T}\coloneqq T-\frac{\tr T}{d-1}\mathds{1}_{d}\,. \label{oppfull}
\end{gather}
 We can rewrite the above in terms of power-traces of the stress-tensor however, the formulæ get significantly more complicated as we travel to higher dimensions, therefore we only provide the first three
\begin{gather}
	\begin{aligned}
	\mathcall{O}_{2}=&\frac{1}{2}\qty[\tr T^{2}-\frac{1}{1}(\tr T)^{2}]=T\overline{T}\,,\\
		\mathcall{O}_{3}=&\frac{1}{2}\qty[\tr T^{2}-\frac{1}{2}(\tr T)^{2}]-\frac{1}{3}\la\qty[\tr T^{3}-\frac{3}{4}\tr T^{2}\tr T+\frac{1}{8}(\tr T)^{3}]\,,\\
			\mathcall{O}_{4}=&\frac{1}{2}\qty[\tr T^{2}-\frac{1}{3}(\tr T)^{2}]-\frac{1}{3}\la\qty[\tr T^{3}-\frac{1}{2}\tr T^{2}\tr T+\frac{1}{18}(\tr T)^{3}]+\\&+\frac{1}{4}\la^{2}\qty[\tr T^{4}-\frac{8}{9}\tr T^{3}\tr T-\frac{1}{2}(\tr T^{2})^{2}+\frac{5}{9}\tr T^{2}(\tr T)^{2}-\frac{11}{162}(\tr T)^{4}]\,.
					\end{aligned} \label{opps}
\end{gather}
One can then easily check that these operators satisfy \eqref{traceflowdd} with \(\Delta_{\la}=-d\) on-shell for the auxiliary background and notice that the higher order contributions could be treated as small \(\lambda\) corrections to the usual \(T^{2}\) operator of \cite{Taylor:2018xcy,Hartman:2018tkw} i.e.
\begin{gather}
	\mathcall{O}_{d}=\frac{1}{2}\qty[\tr T^{2}-\frac{1}{d-1}(\tr T)^{2}]+ \text{finite $\la$ corrections}\,. \label{oppddd}
\end{gather}
Since the leading piece is valid at large \(N\) it is very natural to postulate that our result includes (possibly all) finite \(N\) corrections.\footnote{Assuming that large \(N\) implies small \(\la\) and vice-versa.} However, we must mention that the stress-tensor in the expressions above is always the deformed one, meaning that there is an implicit dependence on \(\la\). This is very important because in order to properly perform the power counting one needs to expand everything and rearrange the terms appropriately.  The situation now becomes very complicated since the leading piece will contribute a term at order \(\la\),  complementing the zeroth order contribution from the first \(\la\) correction. Similarly, at \(\la^{2}\) one gets a piece from the second order expansion of \(T^{2}\) plus a term from the first order expansion of \(T^{3}\) but also the zeroth order part of \(T^{4}\) (for \(d\geq4\)). From this discussion it becomes evident that the actual \(\la\) expansion around zero looks like
\begin{gather}
	\mathcall{O}_{d}=\frac{1}{2}\qty[\tr T_{0}^{2}-\frac{1}{d-1}(\tr T_{0})^{2}]+ \text{infinite $\la$ corrections}\,, \label{oppdd0}
\end{gather}
where \(T_{0}\) is the undeformed stress-tensor which is independent of \(\la\). It is now very easy to confuse \eqref{oppddd} with \eqref{oppdd0}, since they both have the same structure and especially when it comes to comparisons with the holographic approaches \cite{Taylor:2018xcy,Hartman:2018tkw}. The Brown-York stress-tensor there seems to be the deformed one, but the Hubbard-Stratonovich interpretation \cite{Cardy:2018sdv, Hartman:2018tkw} seems to prefer the undeformed stress-tensor. These are conflicting arguments that will require further investigation. 

The field theory analysis we presented is not very illuminating and still does not explain why these corrections are there in the first place. For this reason, we will shift our attention to \eqref{dressmassdfixed} where we have the advantage of working with a relatively simple expression, which also holds for any \(d\). Understanding the dynamics there would then reduce to ``just'' translating them in a field theory language. 

This mass term seems unfamiliar, however we can rewrite it in terms of the corresponding metrics instead by finally utilizing the gauge freedom we have. Our choice will be the symmetric vielbein condition \cite{Hinterbichler:2012cn, Deffayet:2012zc} which simply reads
\begin{gather}
	\tensor{e}{_\mu^a}\tensor{\eta}{_a_b}\tensor{f}{_\n^b}=	\tensor{e}{_\n^a}\tensor{\eta}{_a_b}\tensor{f}{_\m^b}\,, \label{symmviel}
\end{gather}
and as a result \eqref{dressmassdfixed} can now be written as
\begin{gather}
	S_{G}=\frac{1}{\la}\int\dd[d]{x}\sqrt{-\gamma}\,\qty{d-1-\tr\sqrt{\g^{-1}g}+\det\sqrt{\g^{-1}g}}\,.  \label{minimalmassivegr}
\end{gather}
Remarkably, this dressing is well-known amongst the massive gravity community and was initially derived in \cite{Hassan:2011vm,Hassan:2011zd} for \(d=4\). It goes by the name ``ghost-free minimal massive gravity'' and it belongs to the family of dRGT mass terms \cite{deRham:2010kj, deRham:2014zqa}. Very similar constructions also appear in three dimensions, see for example \cite{Bergshoeff:2014pca}. The mass parameter is related to the deformation coupling via \(m^{2}=\#\ka^{2}/\la\) where \(\ka^{2}\) is the usual gravitational coupling constant and \(\#\) some normalization factor. Let us now rewrite the above recalling that the auxiliary background contains all the implicit \(\la\) dependence. We can conveniently parametrize this metric as
\begin{gather}
	\tensor{g}{_\m_\n}=\tensor{\g}{_\m_\n}+2\tensor{h}{_\m_\n}=	\tensor{\g}{_\m_\n}+2\sum_{n=1}^{\infty}\tensor*{h}{^{(n)}_\m_\n}=\tensor{\g}{_\m_\n}+\sum_{n=1}^{\infty}\frac{\la^{n}}{n!}\dv[n]{}{\la}\tensor{g}{_\m_\n}\Bigg{|}_{\lambda\rightarrow 0}\,, \label{expmetric}
\end{gather}
where we essentially factor out the \(\la\rightarrow 0\) behaviour, therefore \eqref{minimalmassivegr} can be equivalently written as
\begin{gather}
	\begin{aligned}
	S_{G}&=\frac{1}{\la}\int\dd[d]{x}\sqrt{-\gamma}\,\sum_{\ka=2}^{d}\text{det} _{\ka}\qty[\sum_{i=1}^{\infty}\binom{\frac{1}{2}}{i}(2\g^{-1}h)^{i}]\\&= \frac{1}{2\la}\int\dd[d]{x}\sqrt{-\gamma}\,\tensor{\g}{^\m^\n}\tensor{\g}{^\al^\be}\qty{\tensor*{h}{^{(1)}_\m_\n}\tensor*{h}{^{(1)}_\al_\be}-\tensor*{h}{^{(1)}_\m_\be}\tensor*{h}{^{(1)}_\n_\al}}+\dots\,,  \label{minimalmassiveexp}
	\end{aligned}
\end{gather}
where \(\text{det}_{\ka}\) is given by \eqref{def detk}. The first thing to notice is that in a \(\la\) expansion, the leading piece is precisely the Fierz-Pauli mass term \cite{Hassan:2011vm,Hassan:2011zd}. Plugging the above in \eqref{sol1} and expanding the seed action in the same way one obtains
\begin{gather}
	S_{\la}=S_{0}+ \frac{1}{2\la}\int\dd[d]{x}\sqrt{-\gamma}\,\tensor{\g}{^\m^\n}\tensor{\g}{^\al^\be}\qty{\tensor*{h}{^{(1)}_\m_\n}\tensor*{h}{^{(1)}_\al_\be}-\tensor*{h}{^{(1)}_\m_\be}\tensor*{h}{^{(1)}_\n_\al}}+\int\dd[d]{x}\sqrt{-\gamma}\,\tensor{{T_{0}}}{^\mu^\nu}\tensor*{h}{^{(1)}_\m_\n}+\dots\,,  \label{minimalmassiveexp2}
\end{gather}
where crucially the undeformed stress-tensor is not an implicit function of \(\la\). This is precisely the construction considered in \cite{Hartman:2018tkw} where the metric \(\tensor*{h}{^{(1)}_\m_\n}\) is understood as the Hubbard-Stratonovich field that needs to be integrated out in a similar way to ours. However, notice that the stress-tensor we consider here is the undeformed one. The solution to the equations of motion is simply
\begin{gather}
\tensor*{h}{^{(1)}_\m_\n}=\la\qty(\tensor{{T_{0}}}{_\mu_\nu}-\frac{\tr T_{0}}{d-1}\tensor{\g}{_\m_\n})\,, \label{solaux}
\end{gather} 
therefore the on-shell action is now
\begin{gather}
	S_{\la}=S_{0}+\frac{\la}{2}\int\dd[d]{x}\sqrt{-\gamma}\, \qty[\tr T_{0}^{2}-\frac{1}{d-1}(\tr T_{0})^{2}]+\dots\,,
\end{gather}
which is very clearly the first order correction to the undeformed action under the deformation \eqref{oppdd0}. One can then keep generating more corrections by integrating out the higher-order auxiliary backgrounds reaching, eventually, the full effect of the deformation i.e. \eqref{oppfull}, covering at the same time the holographically derived \(T^{2}\) part. The conclusion we draw from this analysis is that the random geometries framework seems to be valid only at leading order in the \(\lambda\) expansion if one considers just the contribution from the Fierz-Pauli mass term.

Combining all the information of this subsection, we observe that the \(T^{2}\) operator is related to the linearized Fierz-Pauli dressing (and its derivative corrections) whereas the full \(T^{d}\) deformation comes from its non-linear completion, which enforces the absence of fictitious gravitational degrees of freedom \cite{deRham:2010kj, deRham:2014zqa,Hassan:2011vm,Hassan:2011zd}. Therefore, we finally understand that the higher-order corrections appearing in the deforming operator are necessary since they ensure that spin-2 ghosts are not present in the deformed theory! This can only be seen when we include kinetic terms for the auxiliary background in the mix, which automatically complicates the process of integrating it out. It would be very interesting to further investigate the properties of this gravitational mass term in a more applied setting, so the next subsection will be devoted to that.
\subsection{Dynamical coordinates} 
Coming back to \eqref{dressmassdfixed}, we can now combine it with the undeformed action \(S_{0}\), take a flat space limit and restore diffeomorphism invariance in an exact analogy to the previous section. It is then very natural to postulate that the deformed partition function will be 
\begin{gather}
			\mathcal{Z}_{\la}=\int\frac{\mathcall{D}e\mathcall{D}\mathcal{Y}}{\text{vol}(\text{diff})}\exp{\frac{i}{\la}\int\dd[d]{x}\det(\partial_{\mu}\mathcal{Y}^{a})\qty(d-1-\pdv{x^{\mu}}{\mathcal{Y}^{a}}\tensor{e}{_\mu^a}+\frac{e}{\det(\partial_{\mu}\mathcal{Y}^{a})})}\mathcal{Z}_{0}[e]\,, \label{path4}
\end{gather}
where we have used that \(\tensor{f}{_\m^a}=\partial_{\mu}\mathcal{Y}^{a}\). From here, the first step is to check whether the auxiliary background is also flat, which we do by considering the equations of motion\footnote{Here we define the antisymmetrization of the indices as \(\tensor{A}{_{[\mu}_{\nu]}}\coloneqq\frac{1}{2!}(\tensor{A}{_{\mu}_{\nu}}-\tensor{A}{_{\nu}_{\mu}})\).}  for \(\mathcal{Y}^{a}\) i.e.
\begin{gather}
	\pdv{x^{[\mu}}{\mathcal{Y}^{a}}	\pdv{x^{\nu]}}{\mathcal{Y}^{b}}\partial_{\mu}\tensor{e}{_\n^b}=0\,,
\end{gather}
where one can see are trivially satisfied for \(\tensor{e}{_\mu^a}=\partial_{\mu}\mathcall{Z}^{a}\). Then, we proceed by picking the unitary gauge \cite{Tolley:2019nmm} for \(\tensor{e}{_\m^a}\), that is \(\mathcall{Z}^{a}=x^{a}\), eventually leading to the trivial vielbein as expected. Independently, we can write down the equations of motion \eqref{intout} for \eqref{dressmassdfixed} coupled to \(S_{0}\) or
\begin{gather}
	f\tensor{f}{^{\mu}_{a}}=e \tensor{e}{^{\mu}_{a}}+ e\la \tensor{{T_{0}}}{^{\mu}_{a}}\,, \label{eqmotd}
\end{gather}
and gauge fix \(\tensor{e}{_\mu^a}=\tensor{\de}{_\mu^a}\) only. As a result, an effective change of coordinates interpretation becomes apparent and in order to verify this, we act with \(\partial_{\mu}\) on either side of \eqref{eqmotd} yielding
\begin{gather}
	2f\tensor{f}{^{[\mu}_{a}}\tensor{f}{^{\nu]}_{b}}\partial_{\mu}\tensor{f}{_\n^b}=\la\partial_{\mu}\tensor{{T_{0}}}{^{\mu}_{a}}\,, \label{condition coc}
\end{gather}
therefore on-shell for the fields of \(S_{0}\), these equations hold if \(\tensor{f}{_\m^a}=\partial_{\mu}\mathcal{Y}^{a}\), in agreement with the analysis above. In two dimensions, the left hand side then gives precisely the integrability condition, however this is not true any more in every other higher-dimensional case. The reason behind this behaviour is very simple and goes beyond the specific deformation we consider so let us briefly explain. There are only two conditions that enter this argument, the integrability constraint \(\partial_{[\mu}\partial_{\nu]}\mathcal{Y}^{a}=0\) and the conservation of the undeformed stress-tensor \(\partial_{\mu}\tensor{{T_{0}}}{^{\mu}^{\nu}}=0\). Since both will always be related for any stress-tensor deformation due to \eqref{intout}, we can count the number of equations we get in the end. On one hand, the integrability condition gives \(d \times d(d-1)/2\) relations whereas the conservation of the stress-tensor only \(d\). It is then clear for which (positive) \(d\) these two match, one necessarily needs to fix \(d=2\) or, work with an ``integrated'' condition that sums over the extra \(d(d-1)/2\) degrees of freedom. In our case, this is precisely the job of \(\tensor{f}{^{[\mu}_{a}}\tensor{f}{^{\nu]}_{b}}\) as one can clearly observe from \eqref{condition coc}. 

Following now the standard lore of this approach, we proceed by eliminating all \(\tensor{f}{^\m_a}\) from the deformed action which reads 
\begin{gather}
	S_{\la}=\int\dd[d]{x}f\,\mathcal{L}_{\la}=\frac{1}{\la}\int\dd[d]{x}f\,\qty{d-1-y_{1}+\det Y_{1}}+\int\dd[d]{x}e\,\mathcal{L}_{0}\,,
\end{gather}
via \eqref{eqmotd} and take \(\tensor{e}{_\mu^a}=\tensor{\de}{_\mu^a}\). Remarkably, all gravitational elements have now vanished, leaving behind only the field theory deformation i.e.
\begin{gather}
	S_{\la}=\frac{1}{\la}\int\dd[d]{\mathcal{Y}}\,\qty{d-1-\qty[d-1-\la\qty(\mathcal{L}_{0}-\tr T_{0})]\det(\mathds{1}_{d}+\la T_{0})^{\frac{1}{1-d}}}\,, \label{deffinal}
\end{gather}
and by inverting \eqref{eqmotd}, the accompanying field-dependent change of coordinates\footnote{The Levi-Civita symbol is used below under the convention \(\tensor{\e}{_0_{\cdots}_{d-1}}=\tensor{\e}{^0^{\cdots}^{d-1}}=1\).}
\begin{gather}
	\pdv{\mathcal{Y}^{\nu}}{x^{\mu}}=\dfrac{(\det[ \mathds{1}_{d}+\lambda  T_{0}])^{\frac{2-d}{d-1}}}{(d-1)!}\tensor{\epsilon}{_{\mu}_{\mu_{2}}_{\cdots}_{\mu_{d}}}\tensor{\epsilon}{^{\nu}^{\nu_{2}}^{\cdots}^{\nu_{d}}}( \tensor{\delta}{^{\mu_{2}}_{\nu_{2}}}+\lambda  \tensor{{T_{0}}}{^{\mu_{2}}_{\nu_{2}}})\cdots( \tensor{\delta}{^{\mu_{d}}_{\nu_{d}}}+\lambda  \tensor{{T_{0}}}{^{\mu_{d}}_{\nu_{d}}})\,. \label{coc}
\end{gather} 
 It is crucial to mention that in all three equalities, the deformed and undeformed Lagrangian as well as the undeformed stress-tensor are expressed in the \(x^{\mu}\) coordinate system thus, in order to obtain the actual deformed theory one needs to also re-express
everything in terms of  \(\mathcal{Y}^{a}\) via \eqref{coc}. This is not a trivial task, and it is understandable that all the complexity of this process is hidden in this last step, especially when one wants to check the validity of these expressions. To circumvent this, we will consider the simplest example possible, which avoids entirely this issue and leave more complicated cases for future work. Therefore, we pick a theory consisting of just a potential without any kinetic terms, in other words
\begin{gather}
	\mathcal{L}_{0}=-V\,, \qquad T_{0}=-V\mathds{1}_{d}\,. \label{simple seed}
\end{gather}
 It is then very easy to deduce that the deformed theory is exactly \eqref{deffinal} for this choice. If all these statements are true, the deforming operator must be given by \eqref{oppfull} and a quick check confirms that everything is indeed consistent.
 
 Taking a step back, we observe that our gravitational formalism and deformation leads to a non-trivial generalization of earlier two-dimensional results \cite{Conti:2018tca} and provides a concrete way of generating field-dependent coordinate transformations that yield actions deformed by a stress-tensor operator. One can notice that the source of these findings is yet again the mass term \eqref{dressmassdfixed} and its properties. Before we move to the next subsection where we explore yet another aspect of this approach, we should highlight the importance of \eqref{eom fields}. This equation clearly states that the field saddle points of the deformed and undeformed actions match, which is a strong hint for the existence of a map between the two. But it should be clear from our discussion that this is exactly the content of \eqref{coc}, therefore we conclude that given a seed theory and the solution to its equations of motion on a flat background, one can directly obtain the deformed solution via \eqref{coc}.

\subsection{Non-linear electrodynamics}\label{nonlin}
In this subsection we will take a closer look into the massive gravity description of the Born-Infeld action in four dimensions, generalizing it to the case of a curved background and as a by-product, we provide a concrete way of deforming any Weyl invariant seed theory in four dimensions. We begin from the curved version of the dressing\footnote{We should clarify that the mass term of the expression that follows is not a dRGT term, thus one needs to carefully examine whether ghosts are present in this theory if the Einstein-Hilbert term is also considered.} of \cite{Floss:2023nod}\footnote{We would like to thank Dmitri Sorokin for pointing out the existence of this work.} coupled to Maxwell, which reads
\begin{gather}\label{S1}
	S_{\la}=\frac{1}{4\la}\int\dd[4]{x}\sqrt{-\gamma}\,\qty{4-\sqrt{\gamma^{-1}g}\,\tensor{\gamma}{_{\mu\nu}}\tensor{g}{^{\nu\rho}}\tensor{\gamma}{_{\rho\sigma}}\tensor{g}{^{\sigma\mu}}}-\frac{1}{4}\int\dd[4]{x}\sqrt{-g}\,\tensor{F}{_{\mu\nu}}\tensor{g}{^{\nu\rho}}\tensor{F}{_{\sigma\rho}}\tensor{g}{^{\sigma\mu}}\,.
\end{gather}
 This action has the form of \eqref{sol1}, thus the deformed theory is given by integrating out the auxiliary metric. First, we verify that \eqref{finin} is satisfied. Then, we can proceed by applying our formalism and one solution we find is the following\footnote{Here we should emphasise that if we  were to consider the seed to be a non-abelian gauge theory, the solution for the auxiliary metric would differ from \eqref{sol g BI}, since one must instead treat $\tensor*{F}{^{a}_{\mu\nu}}\tensor*{F}{^{a}_{\rho\sigma}}$ as a single field with four Lorentz indices i.e. $\tensor{X}{_{\mu\nu\rho\sigma}}$. In this case, the resulting deformed action is also valid in the non-abelian limit.}
\begin{gather}\label{sol g BI}
	\tensor{g}{_{\mu\nu}}=\tensor{\gamma}{_{\mu\rho}}\sqrt{\tensor{\delta}{^{\rho}_{\nu}}-\la \tensor{\gamma}{^{\rho\chi}}\tensor{F}{_{\chi\sigma}}\tensor{\gamma}{^{\sigma\tau}}\tensor{F}{_{\tau\nu}}}\,,
\end{gather}
therefore evaluating \eqref{S1} on-shell yields the Born-Infeld action on a curved worldvolume 
\begin{gather}\label{BI}
	S_{\la}=\frac{1}{\la}\int\dd[4]{x}\qty{\sqrt{-\gamma}-\sqrt{-\text{det}(\tensor{\gamma}{_{\mu\nu}}+\sqrt{\la}\tensor{F}{_{\mu\nu}})}}\,.
\end{gather}
At the same time, this model satisfies the flow equation \eqref{floweq} whose operator is given by \cite{Conti:2018jho,Ferko:2019oyv,Babaei-Aghbolagh:2020kjg}
\begin{gather}\label{operator on-shell BI}
	\mathcall{O}=\frac{1}{8}\qty[\tr T^{2}-\frac{1}{2}(\tr T)^{2}]\,,
\end{gather}
hence we aim to examine whether this operator satisfies \eqref{pde} with the gravitational dressing given by \eqref{S1}. In this case the deformed stress-tensor is given by
\begin{gather}\label{stress BI}
	\tensor{T}{^{\mu}_{\nu}}=\frac{1}{\la}\qty(\tensor{\delta}{^{\mu}_{\nu}}-g\gamma^{-1}\tensor{g}{^{\mu\rho}}\tensor{\gamma}{_{\rho\sigma}}\tensor{g}{^{\sigma\tau}}\tensor{\gamma}{_{\tau\nu}})\,,
\end{gather}
and the resulting flow equation, using matrix notation, yields
\begin{gather}\label{flow BI 4d 1}
	\text{det}(g^{-1}\gamma)=\frac{1}{8}(\text{tr} (g^{-1}\gamma)^{2})^{2}-\frac{1}{4}\text{tr} (g^{-1}\gamma)^{4}\,.
\end{gather}
Obviously, this relation does not hold for arbitrary auxiliary metric $\tensor{g}{_{\mu\nu}}$ and even in its current form, it is a highly non-trivial task to check whether \eqref{flow BI 4d 1} holds on-shell for \eqref{sol g BI}. To make the problem tractable, we need to remove the square root dependence, which can be achieved by expressing \eqref{flow BI 4d 1} exclusively in terms of the ``squared'' matrix $(\gamma^{-1} g)^{2}$. After some algebra, we collect the exact $\la$ dependence order-by-order and require each term to vanish. When the dust settles, this yields the following two conditions
\begin{gather}
	\begin{gathered}
		(\text{tr}(\gamma^{-1}F)^{2})^{4}-4\,(\text{tr}(\gamma^{-1}F)^{2})^{2}\text{tr}(\gamma^{-1}F)^{4}-4\,(\text{tr}(\gamma^{-1}F)^{4})^{2}+16\,\text{tr}(\gamma^{-1}F)^{8}=0\,,\\
		(\text{tr}(\gamma^{-1}F)^{2})^{3}-6\,\text{tr}(\gamma^{-1}F)^{2}\text{tr}(\gamma^{-1}F)^{4}+8\,\text{tr}(\gamma^{-1}F)^{6}=0\,,
	\end{gathered}
\end{gather}
which we verified hold for arbitrary field strength and background metric. Hence, we find that the flow equation \eqref{pde} with the operator given by \eqref{operator on-shell BI} is valid only on-shell for the auxiliary metric when the seed theory is Maxwell, as anticipated from our discussion in Section \ref{sec2}. This can mean one of two things; either the mass term of \eqref{S1} or the operator is not universal. In the former case, one should in principle look for solutions  to the flow equation for this operator, which satisfy the usual initial conditions \eqref{finin}. Unfortunately, we could not find any as the resulting equation is a highly non-linear PDE, we therefore proceed by focusing on the latter case.

In other words, we now aim to investigate whether we can construct a universal operator corresponding to the case of arbitrary $S_{0}$ coupled to the dressing of \eqref{S1}. 
To this end,  we should rewrite \eqref{S1} in terms of the $Y_{i}$ matrices as in \eqref{sol1}. After some massaging and using the properties of the elementary symmetric polynomials $\det_{\ka}$ defined in Appendix \ref{conv} we can rewrite the gravitational dressing, using matrix notation, as follows
\begin{gather}\label{G from s11}
	S_{G}=\frac{1}{4\la}\int\dd[4]{x}\sqrt{-\gamma}\,\qty{4-\frac{\text{det}_{3}Y_{1}^{4}}{\text{det} Y_{1}^{3}}}\,.
\end{gather}
Following the massive gravity approach and by fitting an order four polynomial in terms of power-traces of the stress-tensor, we find that the corresponding operator is given by
\begin{gather}\label{OO flow}
	\begin{aligned}
		\mathcall{O}=\frac{16b_{0}-1}{4\la}\tr T+2b_{0}\qty[\tr T^{2}-(\tr T)^{2}]+\frac{2\la b_{0}}{3}\qty[(\tr T)^{3}-3\tr T \tr T^{2}+2\tr T^{3}]&\\
		-\frac{\la^{2}b_{0}}{6}\qty[(\tr T)^{4}-6(\tr T)^{2}\tr T^{2}+3(\tr T^{2})^{2}+8\tr T \tr T^{3}-6\tr T^{4}]&\,,
	\end{aligned}
\end{gather}
where $b_{0}$ is arbitrary. 
This implies that the terms proportional to $b_{0}$ cancel since the left hand side of \eqref{floweq} does not contain any dependence on it, which yields 
\begin{gather}
	\tr T=\la \text{det}_{2}T-\la^{2}\text{det}_{3}T+\la^{3}\text{det}T\,.
\end{gather} 
Let us now compare this trace-flow equation with  \eqref{trflowclass}. First, we notice that there are no curvature nor equation of motion terms. This implies that the seed action $S_{0}$ is necessarily invariant under the Weyl transformations \eqref{rq1}, therefore the gravitational dressing \eqref{G from s11} can only be used to deform theories of this type. By taking this into account together with the fact that the scaling dimension of $\la$ is given by $\Delta_{\la}=-4$, we find that the true operator reads
\begin{gather}\label{gen OO d=4}
	\mathcall{O}_{4}=-\frac{1}{4}\qty[\text{det}_{2}T-\la\text{det}_{3}T+\la^{2}\text{det}T]\,.
\end{gather}
We can now proceed with a few comments. First, we verify that for the Born-Infeld action \eqref{BI}, the operator at hand reduces to \eqref{operator on-shell BI}, however we find that the Dirac-Born-Infeld action is not part of this flow.   Second, we can compare the structure of \eqref{gen OO d=4} with its three-dimensional analogue coming from the D2-brane flow equation, given by \cite{Tsolakidis:2024wut,Blair:2024aqz}
\begin{gather}
	\mathcall{O}_{3}=-\frac{1}{2}\qty[\text{det}_{2}T-\la\text{det}T]\,, \label{3d BI op}
\end{gather}
and so we observe that a potential $d$-dimensional operator has the following structure 
\begin{gather}\label{d-dim O BI}
	\mathcall{O}_{d}=\frac{1}{2^{d-2}\la^{2}}\qty[1-\la \tr T-\text{det}(\mathds{1}_{d}-\la \,T)]\,.
\end{gather}
To this end, it would be interesting to investigate the exact form of the corresponding deformed action for $d\geq 5$ and construct its massive gravity description. Finally, to our surprise, the operator \eqref{d-dim O BI} is exactly of the same form as \eqref{oppfull} under the following transformation for the stress-tensor and simple rescaling of \(\la\)
\begin{gather} \label{transfor}
	\la\longrightarrow2^{1-d/2}\la\,,\qquad T\longrightarrow -2^{d/2-1} \qty(T-\frac{\tr T }{d-1}\mathds{1}_{d})\,.
\end{gather}
This could hint at a universal underlying structure that governs these types of deformations.
\subsection{ Spin zero, and a comment on deformed actions}
In a slightly different spirit, we would like to explore other limits of the Dirac-Born-Infeld action so for this subsection, we will turn off the gauge fields i.e. we will be considering the following deformed theory
\begin{gather}\label{dressing D bosons}
	S_{\la}=\frac{1}{\la}\int\dd[d]{x}\qty{\sqrt{-\gamma}-\sqrt{-\text{det}(\tensor{\gamma}{_\mu_\nu}+\la\tensor{\Phi}{_\mu_\nu})}}\,,
\end{gather}
where \(\tensor{\Phi}{_\mu_\nu}\) is given by \eqref{kin terms of seed action}.  The operator that appears in the corresponding flow equation was found to be \cite{Blair:2024aqz} 
\begin{gather}\label{Operator D bos} \mathcall{O}_{d}=\dfrac{1}{2\la^{2}}\qty[d-2-\la\tr T-(d-2)\qty[\det(\mathds{1}_{d}-\la T)]^{\frac{1}{d-2}}],
\end{gather}
where we once again notice that a trace/determinant structure is present. Interestingly, there does exist another classical operator when one considers the single boson case  \cite{Ferko:2023sps} 
\begin{gather}\label{eqA}
	\mathcall{O}'_{d}=\dfrac{1}{2d}\text{tr}(T^{2})-\dfrac{1}{d^{2}}\tr(T)^{2}-\dfrac{d-2}{2\sqrt{d-1}d^{3/2}}\tr(T)\sqrt{\text{tr}(T^{2})-\dfrac{1}{d}\tr(T)^{2}}\,, 
\end{gather}
which only coincides with \eqref{Operator D bos} in that limit. To make matters worse, we managed to derive a third equivalent operator for this case reading
\begin{gather} \label{eqB}
	\begin{aligned}
		\mathcall{O}''_{d}=\frac{1}{2d}\la\text{tr}(T^{3})-&\frac{1}{2(d-1)}\la\text{tr}(T^{2})\tr(T)+\frac{1}{2d(d-1)}\la\tr(T)^{3}\\&+\frac{d(d-3)+3}{2d(d-1)}\text{tr}(T^{2})-\dfrac{1}{2(d-1)}\tr(T)^{2}\,,
	\end{aligned}
\end{gather}
which enjoys a nicer polynomial form compared to the other two.\footnote{We should note here that we were not able to derive the massive gravity description for all three operators.} In order to make our point transparent, let us restate that we have established the following classical equality for the single boson case
\begin{gather}
	\mathcall{O}_{d}=\mathcall{O}'_{d}=\mathcall{O}''_{d}\,,
\end{gather}
therefore, is it fair to ask which operator would be the one to consider in the end, especially if we want to include quantum effects. The logical choice would be \(\mathcal{O}_{d}\) since this operator is valid also for the sigma model, however this is still not the most general case since we have no gauge field. This means that we might once again be in a position where the current structure of the deformed action allows for other equivalent operators, bringing us in the same awkward spot but a ``level'' above.  In other words, it is not safe to extrapolate by considering a specific example  and then claiming the resulting operator as universal, unless one is absolutely sure that the deformed theory at hand contains all the information about the resulting deformation. One can also notice this for the case of the Born-Infeld action of the previous subsection, the naive operator there is \eqref{operator on-shell BI} and we found the more (but not most) general operator to be \eqref{gen OO d=4}. If we assume that a single D\(p\)-brane is the correct starting point then only the corresponding operator of this model should be considered as a candidate for the higher-dimensional analogue. This of course does not mean that the operators derived by specific limiting cases are not interesting; they can be used as the first check against generalized operators in these limits, exactly like the analysis above. 

Alternatively, one can completely avoid all the aforementioned issues by letting go of the specific deformed theory and attack the problem using our method, which by construction generates deformations that are universal and remains agnostic about the nature of \(S_{0}\). There, we still need to make assumptions regarding the structure of the gravitational mass term and deformation,\footnote{Which should however obey \eqref{pde} and \eqref{finin}.} as the connection between stress-tensor deformations and massive gravity is still not well-understood. 

\subsection{A potential for \((\tr T)^{n}\)} \label{sectr}
In this final subsection we focus on constructing the necessary toolkit for tackling the problem of finding solutions to the flow equation for theories that are deformed via a polynomial of traces of the stress-tensor, in $d$ dimensions. To gain intuition, we start from the two-dimensional case, propose its $d$-dimensional generalization and finally apply our results to a four-dimensional example, extending at the same time already known findings. This being said, let us begin by considering the flow equation of a theory living in two dimensions. Specifically, we aim to find solutions for the following operator
\begin{gather} \label{deff}
	\mathcal{O}=\int\dd[2]{x}f\qty{b_{1}(\tr T)^{2}+b_{2}\dfrac{\tr T}{\lambda}+b_{3}\dfrac{1}{\lambda^{2}}}\,,
\end{gather} 
where \((b_{1}, b_{2}, b_{3})\) are arbitrary dimensionless constants. Applying our formalism, but not \eqref{finin}, we reach the following solution
\begin{gather}\label{def}
	S_{\la}=-\dfrac{1}{16 b_1 \lambda }\int\dd[2]{x}f\{4 b_2+(\tensor{f}{^{\m}_{a}}\tensor{e}{_{\m}^{a}}-2)^2\}+S_{0}\,, 
\end{gather}
where the mass term is not dRGT and we are forced to fix \(b_{3}\) to be
\begin{gather}\label{cond}
	b_{3}=\dfrac{b_{2}(1+b_{2})}{4b_{1}}\,,
\end{gather}
in direct analogy with \eqref{eq:sols in 2d}. For the seed action we will consider a generic model of bosons and fermions in a potential, that is \eqref{seed NLSM}. Following the usual lore, we would proceed by integrating out \(\tensor{e}{_{\m}^{a}}\) via \eqref{intout} however, we find that the system does not have any solutions. This could be problematic for our methodology since the deformed action we are trying to find already exists \cite{Morone:2024ffm}. Taking a closer look at the dynamics of the problem, we realize that the deformation \eqref{deff} is insensitive to Weyl invariant seed actions which can be observed by considering the small \(\la\) expansion of \eqref{def}.  In more detail, we temporarily\footnote{We can undo this for free.} apply \eqref{finin} which renders the \(\la\rightarrow 0\) limit regular leading to \(b_{2}=0=b_{3}\), and consider the first order correction which is then proportional to \((\tr T_{0})^{2}\) according to \eqref{deff}. It is then clear that for a traceless seed theory this deformation has no effect and this will persist to all higher orders. Looking back to Section \ref{sec2}, we realize that the less constraining condition \eqref{condtt} could be used instead, provided we could say something about \(\dd{\tensor{e}{_{\m}^{a}}}/\dd{\la}\). The discussion so far strongly indicates that a potential Weyl symmetry is directly related to our inability to integrate out the auxiliary background so in order to counteract it we will boldly assume that the solutions are always for the form \(\tensor{e}{_{\m}^{a}}=\mathfrak{c}_{\la}\tensor{f}{_{\m}^{a}}\) i.e. a \(\la\)-dependent Weyl rescaling of the reference background. Then, \eqref{condtt} can be recast as
\begin{gather}\label{eq2}
	\tensor{e}{_{\m}^{a}}\dfrac{\delta S_{\la}}{\delta\tensor{e}{_{\m}^{a}}}=0\,, 
\end{gather}
from which \(\mathfrak{c}_{\la}\) can be determined.

Alternatively, if we want to completely fix $\tensor{e}{_{\m}^{a}}$ using an equation that resembles \eqref{intout} we must include correction terms which remove the traceless part of the stress-tensor of the seed action as these should not contribute to the deformation. After some trial and error, the resulting equation then takes the following form 
\begin{gather}\label{eq3}
	\dfrac{\delta S_{\la}}{\delta\tensor{e}{_{\m}^{a}}}-\qty(\dfrac{\delta S_{0}}{\delta\tensor{e}{_{\m}^{a}}}-\dfrac{\tensor{e}{^{\m}_{a}}}{2}	\tensor{e}{_{\n}^{b}}\dfrac{\delta S_{0}}{\delta\tensor{e}{_{\n}^{b}}})=0\,.
\end{gather}
One may easily verify that by multiplying \eqref{eq3} with \(\tensor{e}{_{\m}^{a}}\) one retrieves \eqref{eq2} and by solving the former equation for \eqref{def} and \eqref{seed NLSM} one finds
\begin{gather}\label{sol2d}
	\tensor{e}{_{\m}^{a}}=\dfrac{1-\la b_{1}\tensor{f}{^{\n}_{b}}\tensor{\Psi}{^{b}_{\n}}}{1+4V\lambda b_{1}}\tensor{f}{_{\m}^{a}}\,,
\end{gather}
therefore both approaches are consistent with each other. From here we observe that all the bosonic degrees of freedom completely decouple since they  ``belong'' to the traceless part of the stress-tensor of the seed theory. We may now use this result to retrieve the following deformed action  by plugging it in \eqref{def} leading to
\begin{gather}\label{finaleq}
	S_{\la}=\int\dd[2]{x}f\qty{-\dfrac{b_{2}}{4\la b_{1}}-\dfrac{1}{2}\tensor{\gamma}{^{\mu}^{\nu}}\tensor{\Phi}{_{\mu}_{\nu}}+\dfrac{\la b_{1}(\tensor{f}{^{\mu}_{a}}\tensor{\Psi}{^{a}_{\mu}})^{2}-2\tensor{f}{^{\mu}_{a}}\tensor{\Psi}{^{a}_{\mu}}-4V}{4(1+4V\lambda b_{1})}}\,,
\end{gather}
which up to trivial redefinitions agrees with the deformed action found in \cite{Morone:2024ffm} when the fermions are turned off. As a consistency check we note that \eqref{finaleq} obeys \eqref{deff} when \eqref{cond} is valid. 

Equipped with this example in two dimensions, we can now proceed to its higher-dimensional analogue. 
We begin from the most general case of the family of trace deformations, that is 
\begin{gather}
	\mathcal{O}=\int\dd[d]{x}f\,\mathcall{O}(\la,\tr T)\,, \label{deffk}
\end{gather}
for which it is natural to assume that the deformed action will take the following form 
\begin{gather}
	S_{\la}=S_{G}+S_{0}=\frac{1}{\la}\int\dd[d]{x}f\hspace{0.05cm}K(y_{1})+S_{0}\,. \label{solll}
\end{gather}
Then, the flow equation for this dressing \eqref{pde} yields the higher-dimensional equivalent of \eqref{finin1d} that is
\begin{gather}\label{mastereq}
	K(y_{1})=-\la^{2}\mathcall{O}(\la,[d\,K(y_{1})-y_{1}K(y_{1})']/\la)\,, \qquad K(y_{1})'\coloneqq\dv{K(y_{1})}{y_{1}}\,,
\end{gather} 
which is subject to the usual initial conditions \eqref{finin}. Crucially, the solutions of the equation above provide a complete description for this family of deformations. Moreover notice that we have managed to reduce the problem to a non-linear ODE instead of a PDE, however it is still very complicated even for simple examples. To give a few, first, let us examine the quadratic case to make contact with \eqref{deff} and apply it to a four-dimensional model and then, continue with the scenario where we only pick one term that is \(\sim (\tr T)^{n}\).\footnote{We were able to tame \eqref{mastereq} only for these two cases.} For the quadratic case we find the following solution
\begin{gather}
	K(y_{1})=-\dfrac{1}{2d^{2} b_{1}}\qty[2\qty(\qty[\frac{y_{1}}{d}]^{d/2}-1)^{2}+db_{2}]\,, \qquad b_{3}=\dfrac{b_{2}(2+db_{2})}{4db_{1}}\,. \label{2dsol}
\end{gather}
At $d=2$ we remark that we recover the gravitational dressing of \eqref{def}, as required/expected, and if \eqref{finin} is applied, once again \(b_{2}=0\) in any \(d\). It is then also easy to check that for \(d=1\) one retrieves the kernel of \eqref{path3} for \(b_{1}=-1\).

Given this solution let us further check its validity with a concrete example in four dimensions. Specifically, let us consider a sigma model of spin-0 and spin-1 bosons interacting with an arbitrary potential, that is\footnote{For \(d>2\) dimensions fermions do not behave the same way as for $d=2$, as the spin connection does not cancel, and hence we refrain from including them in the below analysis.} 
\begin{gather}
	S_{0}=\int\dd[4]{x}e\,\qty{-\dfrac{1}{4}\tensor{F}{_{\mu\nu}}\tensor{g}{^{\mu\rho}}\tensor{F}{_{\rho\la}}\tensor{g}{^{\la\nu}}-\dfrac{1}{2}\tensor{g}{^{\mu}^{\nu}}\tensor{\Phi}{_{\mu}_{\nu}}-V}\,. \label{seeeed}
\end{gather}
Again we must supplement \eqref{intout} with the necessary ``corrections'' to  the traceless part of the stress-tensor of \eqref{seeeed}. Then the equations of motion for the complete action take the following form
\begin{gather}\label{eq4}
	\dfrac{\delta S_{\la}}{\delta\tensor{e}{_{\m}^{a}}}-\qty(\dfrac{\delta S_{0}}{\delta\tensor{e}{_{\m}^{a}}}-\dfrac{\tensor{e}{^{\m}_{a}}}{4}	\tensor{e}{_{\n}^{b}}\dfrac{\delta S_{0}}{\delta\tensor{e}{_{\n}^{b}}})=0\,, 
\end{gather}
directly mirroring \eqref{eq3} of the two-dimensional example. Now, we may insert the seed action into \eqref{solll} and use \eqref{2dsol} together with \eqref{eq4} to generate a solution for the vierbein i.e.
\begin{gather}\label{sol4d}
	\tensor{e}{_{\m}^{a}}=\sqrt{\dfrac{1-4\tensor{\g}{^{\mu}^{\nu}}\tensor{\Phi}{_{\mu}_{\nu}}\lambda b_{1}}{1+16V\lambda b_{1}}}\tensor{f}{_{\m}^{a}}\,,
\end{gather}
which is a \(\la\)-dependent Weyl transformation, verifying \eqref{eq2} and the arguments surrounding it.  Notice that the spin-1 fields do not contribute to the solution, as the corresponding stress-tensor is traceless in four dimensions, in direct analogy with the bosons in two dimensions. Then, plugging this solution in \eqref{solll} we reach 
\begin{gather}\label{sol4dd}
	S_{\la}=\int\dd[4]{x}f\qty{-\frac{b_{2}}{8\lambda b_{1}}-\dfrac{1}{4}\tensor{F}{_{\mu}_{\nu}}\tensor{\g}{^{\mu}^{\rho}}\tensor{F}{_{\rho}_{\la}}\tensor{\g}{^{\la}^{\nu}}-\frac{\tensor{\g}{^{\mu}^{\nu}}\tensor{\Phi}{_{\mu}_{\nu}}-2\lambda b_{1}(\tensor{\g}{^{\mu}^{\nu}}\tensor{\Phi}{_{\mu}_{\nu}})^{2}+2V}{2(1+16V\lambda b_{1})}}\,,
\end{gather}
which, up to appropriate rescalings, agrees with the deformed action previously derived in \cite{Morone:2024ffm}. Moreover, we notice that the above deformed action satisfies the initial flow equation \eqref{deff} we aimed to solve, when $b_{3}$ is given by \eqref{2dsol}. Another observation we can make by comparing  \eqref{finaleq} and \eqref{sol4dd} is that spin-0 and spin-1/2 fields in \(d=2\) behave in the same way as spin-1 and spin-0 fields do in \(d=4\).
Finally, looking back at \eqref{eq3} and \eqref{eq4}, it is clear that the equations of motion in arbitrary dimensions now take the following form
\begin{gather}\label{eqd}
	\dfrac{\delta S_{\la}}{\delta\tensor{e}{_{\m}^{a}}}-\qty(\dfrac{\delta S_{0}}{\delta\tensor{e}{_{\m}^{a}}}-\dfrac{\tensor{e}{^{\m}_{a}}}{d}	\tensor{e}{_{\n}^{b}}\dfrac{\delta S_{0}}{\delta\tensor{e}{_{\n}^{b}}})=0\,. 
\end{gather}
Combining these relations with \eqref{mastereq} allows one to fully determine the solution for the auxiliary vielbein.

Finally, we are now in a position where we could tackle a more complicated operator. As mentioned before, consider now the following deformation
\begin{gather}
	\mathcal{O}=\int\dd[d]{x}f\lambda^{n-2}b_{1}(\tr T)^{n}\,,
\end{gather}
where \(b_{1}\) is an arbitrary dimensionless constant and \(n>1\). For this choice \eqref{mastereq} becomes
\begin{gather}\label{newmaster}
	K(y_{1})=-b_{1}(d\,K(y_{1})-y_{1}K(y_{1})')^{n}\,,
\end{gather} 
therefore given \eqref{finin}, the solution reads
\begin{gather} \label{grav1}
	K(y_{1})=-\qty[\dfrac{\qty(y_{1}/d)^{d(n-1)/n}-1}{db_{1}^{1/n}}]^{n/(n-1)}\,.
\end{gather}
Similarly as before, combining the above with \eqref{eqd} we have everything we need to determine the solution to the auxiliary vielbein and obtain a deformed action. One can now easily verify that for \(n=2\) we obtain \eqref{2dsol} with \(b_{2}=0\), but also for \(d=2\) and \eqref{seed NLSM} without fermions the resulting theory matches precisely \cite{Ran:2024vgl} in the single boson limit. In any case, we must note however, that the gravitational structure at hand is highly non-linear and does not resemble or share any properties with dRGT gravity.
\section{Summary and discussion} \label{disc}
In this paper we showed how the massive gravity formulation of stress-tensor deformations in arbitrary dimensions can lead to universal results as well as techniques, which can be used to tackle different problems. To demonstrate the power of this approach, we started exploring two dimensions and used the tool of perturbation theory. As a first consistency check, we show how one can recover $T\overline{T}$ enhanced with $\tr T + \Lambda_{2}$, proving that this extension is not unique to the massive  gravity approach but holds in general. In addition, by studying the deformed action of two different models that feature no fermions but are characterized by a non-trivial potential, we perform a perturbative expansion in a suitably chosen scalar parameter, to find an infinite extension of Schwarz's list \cite{Schwarz1873}. More concretely, for the first case we retrieve \eqref{sum hypers from NLSM} in accordance with \cite{Cavaglia:2016oda, Bonelli:2018kik} and in the second we find that the following Taylor expansion holds
\begin{gather}\label{final1}
	\sum_{n=0}^{\infty}\frac{3^{n-5/2}(-y)^{n}}{\Gamma(1+n)}\tpFq{5}{4}{\frac{1}{2},\,\,\frac{3}{4},\,\,1,\,\,1,\,\,\frac{5}{4}}{2,\frac{3-n}{3},\frac{4-n}{3},\frac{5-n}{3}}{\tfrac{64}{27}z}=\frac{(1-y)^{2}}{4\pi \mathcal{H}^{2}}-\frac{(\mathcal{H}-1)^{2}}{\pi z(1-y)}\,,
\end{gather}
where we defined $\mathcal{H}\coloneqq H_{1,+}$, which is given by \eqref{Hsols} and \eqref{x, P and Fsq defs} in the limit  $ \tensor{\hat f}{_{\mu}^{a}}\rightarrow \tensor{ f}{_{\mu}^{a}}$. A natural next step would be to examine whether other deformed actions also yield similar insights, we hope to report on this in the near future. Crucially, while our methods are anchored on the massive gravity approach, a by-product of our analysis is the formulation of an inverse algorithm, initiating the study of the algebraic properties of special functions directly through the $T\overline{T}$ programme. In particular, we argued that this approach can potentially be applied to other special functions, and given a suitable scalar parameter to perform the perturbative expansion in, it could determine whether the function admits an underlying algebraic structure or not. 

Switching gears, we clarified the role of the trace-flow equation,\footnote{We achieve this in any dimension.} which is secretly a consequence of the local RG flow \cite{Osborn:1991gm},  and highlighted the importance of Weyl symmetry for the underlying theory. We concluded our low-dimensional analysis performing multiple checks in one dimension, achieving among other things a precise match with the gravitational dressing proposed in \cite{Gross:2019ach}. Notably, an essential tool in establishing this were the universal initial conditions for stress-tensor deformations in arbitrary dimensions \eqref{finin}, that we derived in Section \ref{sec2} using the massive-gravity approach. Furthermore, we verified that these initial conditions hold in all examples that were analysed in this work, thereby confirming their robustness. 

Next, we analysed the additional terms appearing in our higher-dimensional analogue of the $T\overline{T}$
operator. Specifically, starting from the general mass term in $d$ dimensions \eqref{dressmassd}, applying \eqref{finin} and combining it with the flow equation given by \eqref{pde}, we find that the resulting operator is given by 
\begin{gather}\label{final2}
	\mathcall{O}_{d}=\frac{1}{\la^{2}}\qty[1+\la\tr\mathcal{T}-\det(\mathds{1}_{d}+\la\mathcal{T})]\,, \qquad \mathcal{T}\coloneqq T-\frac{\tr T}{d-1}\mathds{1}_{d}\,. 
\end{gather}
Subsequently, by expressing this relation in terms of power-traces of the stress-tensor, we find that it can be written as an order \(d\) polynomial in $\la$, that is
\begin{gather}\label{final3}
	\mathcall{O}_{d}=\frac{1}{2}\qty[\tr T^{2}-\frac{1}{d-1}(\tr T)^{2}]+ \text{finite $\la$ corrections}\,.
\end{gather}
Comparing this with the earlier holographic studies of \cite{deHaro:2000vlm,Balasubramanian:1999re,Taylor:2018xcy,Hartman:2018tkw} we find that the zeroth order term exactly reproduces the $T^{2}$ operator. Moreover, the fact that the large $N$ limit yields precisely this latter operator, implies that one can interpret \eqref{final3} as an expansion in $1/N$, suggesting that the additional terms initially found in \cite{Tsolakidis:2024wut} may capture all $1/N$ corrections. From the massive gravity perspective, when implementing the symmetric vielbein condition as our gauge choice, we find that the corresponding mass term takes the following form
\begin{gather} \label{final4}
	S_{G}=\frac{1}{\la}\int\dd[d]{x}\sqrt{-\gamma}\,\qty{d-1-\tr\sqrt{\g^{-1}g}+\det\sqrt{\g^{-1}g}}\,,
\end{gather}
which is remarkably the higher-dimensional version of the famous four-dimensional ghost-free minimal massive gravity action \cite{Hassan:2011vm,Hassan:2011zd, deRham:2010kj,deRham:2014zqa}. From there we found that the leading order $T^{2}$ operator of \eqref{final3} is directly related to the linearized Fierz-Pauli dressing, recovering at the same time the Hubbard-Stratonovich interpretation of \cite{Cardy:2018sdv, Hartman:2018tkw}. 
This implies that all subleading terms essentially entail its non-linear completion, and as a result ensures the absence of spin-2 ghosts. Said differently, we show that the existence of the higher order terms is in this sense imperative for the consistent formulation of a gravitational theory of this type. We should however mention that we do not yet have a clear interpretation of these terms from a field theory point of view. It would be most interesting to further investigate this.

Furthermore, we studied the flat space limit of this gravitational dressing with the hope of uncovering a dynamical coordinates perspective. Surprisingly, we were able to derive a deformed action that lacks gravitational elements and is solely expressed in terms of the undeformed field theory data $\La_{0}$ and $T_{0}$. Explicitly, we have
\begin{gather}
	S_{\la}=\frac{1}{\la}\int\dd[d]{\mathcal{Y}}\,\qty{d-1-\qty[d-1-\la\qty(\mathcal{L}_{0}-\tr T_{0})]\det(\mathds{1}_{d}+\la T_{0})^{\frac{1}{1-d}}}\,, 
\end{gather}
accompanied with the following field-dependent change of coordinates
\begin{gather}\label{final5}
	\pdv{\mathcal{Y}^{\nu}}{x^{\mu}}=\dfrac{(\det[ \mathds{1}_{d}+\lambda  T_{0}])^{\frac{2-d}{d-1}}}{(d-1)!}\tensor{\epsilon}{_{\mu}_{\mu_{2}}_{\cdots}_{\mu_{d}}}\tensor{\epsilon}{^{\nu}^{\nu_{2}}^{\cdots}^{\nu_{d}}}( \tensor{\delta}{^{\mu_{2}}_{\nu_{2}}}+\lambda  \tensor{{T_{0}}}{^{\mu_{2}}_{\nu_{2}}})\cdots( \tensor{\delta}{^{\mu_{d}}_{\nu_{d}}}+\lambda  \tensor{{T_{0}}}{^{\mu_{d}}_{\nu_{d}}})\,. 
\end{gather} 
Staring at these for while, one realizes that this is a $d$-dimensional counterpart of the two-dimensional dynamical coordinates formulation of \(T\overline{T}\) \cite{Conti:2018tca}. To further substantiate this, we also show that the saddles of the deformed and undeformed theories match \eqref{eom fields}, suggesting that the equation above serves as the map between the two. As mentioned previously, verifying that this coordinate transformation indeed gives the correct deformed action, proves to be a highly non-trivial task. However, we have checked our results by limiting ourselves to the case where the seed theory only contains a potential and no kinetic terms. Nonetheless, we aim to further test our findings with more complicated examples and report on this in future work. 

Focussing for a moment on $d=4$ and the Born-Infeld action, we studied the curved background version of the gravitational dressing first introduced in \cite{Floss:2023nod}, coupled to Maxwell theory. As a result, we find an operator that can be used to deform any Weyl invariant seed theory in four dimensions. By comparing this deformation to the one corresponding to the D2-brane action, we observe a common structure and propose that its $d$-dimensional generalization reads
\begin{gather}\label{final6}
	\mathcall{O}_{d}=\frac{1}{2^{d-2}\la^{2}}\qty[1-\la \tr T-\text{det}(\mathds{1}_{d}-\la \,T)]\,.
\end{gather}
From there we observe that \eqref{final2} and \eqref{final6} are related by an appropriate transformation  \eqref{transfor} of the stress-tensor $T$ and the deformation parameter $\la$. Given this connection between these seemingly distinct operators it would be interesting to understand whether there also exists a similar equivalence between their gravitational counterparts. 

Continuing our $d$-dimensional journey, we comment on the different operators that can be obtained by investigating the flow equation when the deformed theory is given by the Dirac-Born-Infeld action in curved spacetime with the gauge fields turned off. Presently, we are unable to reconstruct the massive gravity description of these different operators, but it would nonetheless be interesting to achieve this. We expect that this requires implementing a suitable ansatz for the dressing in terms of the $Y_{i}$ matrices that differs from the ones studied in this work.

Finally, we studied a certain family of stress-tensor deformations in $d$ dimensions, whose operators are exclusively expressed as a polynomial of $\tr T$. Gaining intuition from specific seed actions, we find that such deformations have no effect when we consider Weyl invariant seed theories. Consequently, our analysis suggests that the background is also invariant under Weyl, hence the equations of motion of the auxiliary vielbein $\tensor{e}{_{\mu}^{a}}$ must be supplemented with appropriate ``correction'' terms \eqref{eqd}. In addition, we derive the corresponding gravitational dressing for certain limiting cases of the respective operator, which interestingly do not resemble any known mass term. Equipped with these results, we now have the necessary toolkit to tackle the problem of finding a deformed action for any operator of this family.

Apart from the open problems discussed above, it would be interesting to further understand how one would proceed to deform a purely gravitational theory or more generally, a seed action that contains a dynamical background. Some work has already been done in two \cite{Bonelli:2018kik} and four \cite{Morone:2024ffm} dimensions, with the motivation coming from dilaton and \(f(R)\) gravities respectively. We have deliberately postponed this discussion to this section since it is clear from our point of view that integrating out an auxiliary vielbein for some stress-tensor deformation in this case, can be equivalent to a full-blown GR problem i.e. the solutions (if they exist) are to be given by a set of non-linear PDEs. That is, if one is working in the second order formalism where the spin connection is usually picked to be the torsion-less Levi-Civita, which is the case for \cite{Bonelli:2018kik}. There, the authors were only able to find a leading order approximation of the deformed theory containing non-trivial higher-derivative terms of the matter fields. We can now identify the source of this complexity with the non-linear and differential nature of the resulting equations of motion for the auxiliary background. 

On the other hand, one can postpone this issue if the first order formalism is employed instead \cite{Morone:2024ffm}. The spin connection can now be treated as an independent quantity and as a result, any \(f(R)\) kinetic term will only contain an algebraic dependence on the background. This automatically means that finding the saddle point for the background now reduces to a solution of an algebraic equation, which is in principle, solvable. Assuming that this can be done, it is then understood that the deformed action will contain an undetermined spin connection that one needs to integrate out. We can then identify two interesting and potentially problematic issues that may arise. The first would be that it is not really clear if these two steps necessarily commute, considering that this type of situation is identical to what happens with fermions on a curved background, see \cite{Freedman:2012zz} for a very nice explanation. The second and more serious concern is that, at the end of the day, one still needs to integrate out the spin connection in order to reach the physical theory, which may not be feasible considering the intricate structure at hand. Nevertheless, we are very curious to find out if there is a concrete way of including dynamical gravity in the mix, as the current understanding of these deformations seems to be restricted to ordinary matter. 
\section*{Acknowledgments}
We would like to thank Chris D. A. Blair, Kyung-Sun Lee, Tommaso Morone, Niels A. Obers, Matthew M. Roberts, Roberto Tateo, L\'arus Thorlacius, Ziqi Yan, and Junggi Yoon for interesting conversations. We are especially grateful to Friðrik Freyr Gautason, Valentina Giangreco M. Puletti and Abhiram Kidambi for extensive discussions and comments on the manuscript.  
This work was supported by the Icelandic Research Fund under grant 228952-053,  the Eimskip Fund of the University of Iceland, the Aðalsteinn Kristjánsson Memorial Fund and the COST Action CA22113.  AN is grateful to the University of Southampton and ET is grateful to the Niels Bohr Institute and  NORDITA, for the kind hospitality during the early stages of this work. AN and ET would like to thank KU Leuven and the University of Oviedo for their warm hospitality during the completion of parts of this work.

\appendix
\section{The \(\kappa\)-determinant} \label{conv}
Here we define the \(\kappa\)-determinant as the elementary symmetric polynomials of a given $d\times d$ matrix $A$ \cite{Hinterbichler:2012cn}
\begin{gather}\label{def detk}
	\text{det}_{\ka}(A)\coloneqq \frac{(-1)^{\ka}}{\ka!}B_{\ka}(-\tr A,-1!\tr A^{2},-2! \tr A^{3},\dots,-(\ka-1)!\tr A^{\ka})\,,
\end{gather}
where $B_{\ka}$ is the complete exponential Bell polynomial. Intuitively, one may think about this expression as a generalization of the determinant of $A$, but by treating the matrix as if it was $\ka\times\ka$ instead of $d\times d$-dimensional. Hence, it satisfies the following properties
\begin{gather}\label{properties detk}
	\begin{aligned}
		\text{det}_{\ka}(\la A)=\la^{\ka}\text{det}_{\ka}(A)\,, \qquad \text{det}_{\ka}(\mathds{1}_{d})=\binom{d}{\ka}\,,\qquad \text{det}_{\ka}(A^{\text{T}})&=\text{det}_{\ka}(A)\,, \\
		\text{det}_{0}(A)\coloneqq 1 \,,\qquad 	\text{det}_{1}(A)\coloneqq\tr(A)\,, \qquad	\text{det}_{d}(A)&\coloneqq \det(A)\,,
	\end{aligned}
\end{gather}
as well as
\begin{gather}\label{det k inverse}
	\text{det}_{\ka}(A^{-1})=\frac{\text{det}_{d-\ka}(A)}{\det( A)}\,,
\end{gather}
where $\la$ is a constant.
\renewcommand{\refname}{\centering{References}}  
\addcontentsline{toc}{section}{\protect\numberline{}\protect\hspace*{-\cftsecnumwidth}References}
\bibliographystyle{bibstyle}
\bibliography{sample}
\end{document}